\documentclass[traditabstract]{aa}
 \usepackage{txfonts}
 \usepackage{natbib}
 \usepackage{graphicx,multirow}
 \usepackage{color}
 \newcommand{\Fe}{\mbox{\ion{Fe}{ii}}}
 \newcommand{\Ox}{\mbox{[\ion{O}{iii}]}}
 \newcommand{\Ni}{\mbox{[\ion{N}{ii}]}}
 \newcommand{\Su}{\mbox{[\ion{S}{ii}]}}
 \newcommand{\Hb}{\mbox{H$\beta$}}
 \newcommand{\Ha}{\mbox{H$\alpha$}}
 \bibpunct{(}{)}{;}{a}{}{,}

\begin{document}

\titlerunning{Mapping the ionised gas around the luminous QSO HE~1029$-$1401}
\title{Mapping the ionised gas around the luminous QSO HE~1029$-$1401: Evidence for minor merger events?
\thanks{Based on observations made with VIMOS integral field spectrograph mounted on the Melipal VLT telescope
at ESO-Paranal Observatory (programme 072.B-0550A; PI: K. Jahnke)}}

\author{B. Husemann\inst{1}
  \and S. F. S\'anchez\inst{2,3,4} 
  \and L. Wisotzki\inst{1}
  \and K. Jahnke\inst{5}
  \and D. Kupko\inst{1}
   \and D. Nugroho\inst{5} 
   \and M. Schramm\inst{1,6}
}

\institute{Astrophysikalisches Institut Potsdam, An der Sternwarte 16, 14482 Potsdam, Germany  
\and 
Centro de Estudios de F\'\i sica del Cosmos de Aragon (CEFCA), C/General Pizarro 1, E-41001 Teruel, Spain
\and
Fundaci\'on Agencia Aragonesa para la Investigaci\'on y el Desarrollo (ARAID)
\and 
Centro Astron\'omico Hispano-Alem\'an, Calar Alto,(CSIC-MPG), C/Jes\'us Durb\'an Rem\'on 2-2, E-04004 Almeria, Spain
\and
Max-Planck-Institut f\"ur Astronomie, K\"onigsstuhl 17, D-69117 Heidelberg, Germany
\and
Department of Astronomy, Kyoto University, Kyoto 606-8502, Japan 
}

\date{}
\abstract{We present VIMOS integral field spectroscopy of the brightest radio-quiet QSO on the southern sky HE~1029$-$1401 at a redshift of $z=0.086$. Standard decomposition techniques for broad-band imaging are extended to integral field data in order to deblend the QSO and host emission.  We perform a tentative analysis of the stellar continuum finding a young stellar population ($<100$\,Myr) or a featureless continuum embedded in an old stellar population (10\,Gyr) typical for a massive elliptical galaxy. The stellar velocity dispersion of $\sigma_*=320\pm90$\,km/s and the estimated black hole mass $\log(M_\mathrm{BH}/M_{\sun})=8.7\pm0.3$ are consistent with the local $M_\mathrm{BH}$--$\sigma_*$ relation within the errors. For the first time we map the two-dimensional ionised gas distribution and the gas velocity field around HE~1029$-$1401. While the stellar host morphology is purely elliptical we find a highly structured distribution of ionised gas out to 16\,kpc from the QSO. The gas is highly ionised solely by the QSO radiation and has a significantly lower metallicity than would be expected for the stellar mass of the host, indicating an external origin of the gas most likely due to minor mergers. We find a rotating gas disc around the QSO and a dispersion-dominated non-rotating gas component within the central 3\,kpc. At larger distances the velocity field is heavily disturbed, which could be interpreted as another signature of past minor merger events. Alternatively, the arc-like structure seen in the ionised gas might also be indicative of a large-scale expanding bubble, centred on and possibly driven by the active nucleus.}
\maketitle

\section{Introduction}
Since the underlying host galaxies of several quasi-stellar objects (QSOs) were resolved by \citet{Kristian:1973}, QSO host galaxies have been extensively studied to understand their properties and their relevance for the evolution of the overall galaxy population over cosmic time. Most of our knowledge of QSO host galaxies comes from numerous ground and space-based imaging studies \citep[e.g.][]{Malkan:1984,Smith:1986,McLeod:1994,Bahcall:1997,Dunlop:2003,Sanchez:2003}, which provide basic information on the host galaxies like the morphology or the luminosity. Narrow-band images centred on luminous emission lines have been used obtained to infer the distribution and size of the ionised gas surrounding the QSO \citep[e.g.][]{Stockton:1987,Mulchaey:1996,Falcke:1998,Bennert:2002,Schmitt:2003a}. However, optical and/or near-infrared spectroscopy is required for a more detailed investigation of the properties of the stellar population and the ionised gas. 

\citet{Boroson:1982} were the first to detect strong Balmer absorption lines in the off-nuclear spectrum of the  QSO 3C~48 and highly ionised gas was already found long ago in off-nuclear spectra of many QSOs \citep[e.g.][]{Wampler:1975,Stockton:1976,Boroson:1984,Boroson:1985} indicating large amount of gas being photoionised  by the QSO radiation. Unfortunately, spectroscopy of QSO hosts suffers invariably from the contamination of the spectrum by the emission of the active galactic nucleus (AGN) even when observed some arcseconds away from the nucleus. Various spectroscopic studies circumvented this problem by looking at obscured (type 2) AGN \citep[e.g.][]{Gonzalez-Delgado:2001,Kauffmann:2003,CidFernandes:2004,Bennert:2006a,Stoklasova:2009}, which are thought to have the same properties as unobscured (type 1) AGN in the framework of the unification model of AGN \citep{Antonucci:1993,Urry:1995}. They often find a young stellar population in the AGN host galaxies, in particular for the most luminous AGN, suggesting a connection between star formation and nuclear activity in agreement with photometric studies of type 1 QSOs at all redshifts \citep{Jahnke:2003,Jahnke:2004b,Jahnke:2004c,Sanchez:2004b}.

\citet{Letawe:2007} and \citet{Jahnke:2007}, hereafter Let07 and Jah07 respectively, developed two different methods to decompose the AGN and host spectra in long-slit observations and applied them to a sample of luminous QSOs. However, long-slit spectroscopy covers only a small part of the host on the chosen position angle and slit width. Only integral field unit (IFU) observations allow to study the properties of the entire host galaxy in an unbiased way. Dedicated AGN-host decomposition techniques have already been developed and were successfully applied to IFU data of luminous QSOs at low and high redshift \citep{Sanchez:2004a,Christensen:2006,Husemann:2008}.

In this paper, we report on optical IFU spectroscopy of the luminous radio-quiet QSO \object{HE 1029$-$1401} with the VIMOS IFU at the Very Large Telescope (VLT). HE~1029$-$1401 was discovered by the Hamburg-ESO survey \citep[HES;][]{Wisotzki:1991,Wisotzki:2000} to be the brightest QSOs in the southern hemisphere with an apparent magnitude of $V=13.7$ at a redshift of only $z=0.086$. 

Observations with the Hubble Space Telescope revealed a bright elliptical (E1) host galaxy \citep{Bahcall:1997} with an effective radius of $r_\mathrm{e}=3.2\arcsec$ (5.1\,kpc). \citet{Jahnke:2004b} found from multi-colour imaging that the host colours are significantly bluer than inactive galaxies at the same luminosity and inferred an intermediate-age stellar population of 0.7\,Gyr (luminosity weighted) by modelling the spectral Energy distribution (SED) with template Single Stellar Populations (SPPs).

Spatially resolved on-nuclear longslit spectroscopy for this object was presented by Let07 and Jah07. Both studies consistently found highly ionised gas in the host galaxy on kpc scales. Jah07 found that the stellar component is non-rotating, but  the velocity curve of the ionised gas indicated the presence of a rotating gas disc on kpc scales around the nucleus. Let07 argued instead that the velocity curve does not fit with a pure rotational velocity field based on inclination arguments. 

The main focus of this paper is to present an in-depth analysis of the ionised gas in the host galaxy of HE~1029$-$1401, but we also perform a tentative analysis of the stellar continuum. Throughout the paper we assume a cosmological model with $H_0=70$\,km/s, $\Omega_\mathrm{m}=0.3$, and $\Omega_\Lambda=0.7$. This corresponds to a physical scale of $1.6$\,kpc/\arcsec\ at $z=0.086$.

\section{Morphology of the host galaxy of HE~1029$-$1401}

\begin{figure}
\resizebox{\hsize}{!}{\includegraphics[clip]{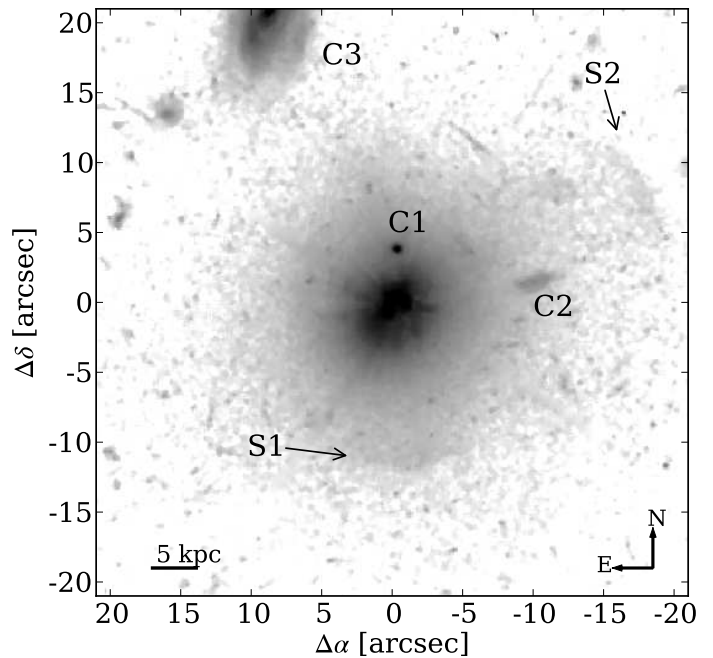}}
\caption{Nucleus-subtracted HST F606W broad-band image of the host galaxy of HE~1029$-$1401. Residuals of the diffraction spikes have been taken out by interpolation and the image is smoothed by a median filter (boxsize of 4 pixels) for display purposes. Three apparent nearby companion galaxies are labelled as C1, C2 and C3 of which only C3 is confirmed to be at a different redshift \citep{Wisotzki:1994}. The location of the two faint shells reported by \citet{Bahcall:1997} are highlighted with arrows.}
\label{fig:HST_image}
\end{figure}

A high resolution HST image of HE~1029$-$1401 in the F606W band was published by \citet{Bahcall:1997}. 
We retrieved the archival HST image from the Hubble Legacy Archive\footnote{http://hla.stsci.edu/} in order to re-analyse the image. An analytic point spread function (PSF) for the observations was created with the TinyTim software package \citep{Krist:1995}, because no sufficiently bright star was covered by the HST observation. We used GALFIT \citep{Peng:2002} to decompose the host and QSO contributions assuming a  de Vaucouleurs light profile \citep{deVaucouleurs:1948} for the host plus a point source for the nucleus. The effective radius and ellipticity of our best-fitting model was $r_\mathrm{e}=3.5\arcsec$ (5.6\,kpc)  and $e=0.1$, respectively, which are consistent with the parameters derived by \citet{Bahcall:1997}. 

The nucleus-subtracted image of HE~1029$-$1401 is shown in Fig.~\ref{fig:HST_image}. A strong negative residual is still visible at the QSO position due to the saturated QSO core. We cleaned the image of residuals from the usual diffraction spikes by replacing each affected pixel with the median value of the unaffected pixels within a radius of 6 pixels. \citet{Wisotzki:1994} obtained redshifts for 13 galaxies in the field finding 4 physical companions which indicate that the HE~1029$-$1401 is part of a loose group of galaxies. None of these confirmed companions were covered with our IFU observations. Neither C1 nor C2 have redshift information available, but C3 was found to be at a different redshift than HE~1029$-$1401 \citep{Wisotzki:1994}.

\citet{Bahcall:1997} also reported the detection of two faint shells located roughly 11\arcsec\ and 19\arcsec\ away from the nucleus. These shells are hardly visible in Fig.~\ref{fig:HST_image}, but we highlighted their position by arrows for guidance and labelled them as S1 and S2, respectively. Note that the shell S2 is outside the field-of-view of our IFU observations.

We also performed an isophotal analysis of the host using the \emph{ellipse} task of IRAF\footnote{IRAF is distributed by the National Optical Astronomy Observatories,
    which are operated by the Association of Universities for Research
    in Astronomy, Inc., under cooperative agreement with the National
    Science Foundation.} 
\citep{Tody:1993} after masking the 2 closest apparent companions. As the image is strongly affected by decomposition residuals over the central 2\arcsec, we only took into account isophotes within a range of 2\arcsec--10\arcsec\ in the semi-major axis. We found that the ellipticity of the host is slightly decreasing outwards from 0.2--0.1 and that the $a_4/a$ Fourier coefficient is consistent with 0. Thus, the deviation from pure elliptical isophotes is marginal.

\section{IFU Observations and Data Reduction}\label{sect:obs_reduct}
We employed the IFU mode of the VIMOS instrument \citep{LeFevre:2003} on the VLT to perform optical 3-D spectroscopy of HE~1029$-$1401. The observations were conducted on the 18th and 24th of December 2003 at a seeing of $\sim 1.1\arcsec$, with the high-resolution blue and orange grisms covering the spectral regions around H$\beta$ and H$\alpha$ with a spectral resolution of $R\sim2600$.
We chose a spatial resolution of 0.66\arcsec/fibre resulting in a field-of-view of 27\arcsec$\times$27\arcsec\ that matches well with the angular size of the QSO host galaxy. The exposure times were split into $3\times 300$s for the blue and $6\times 450$s for the orange grism. A dither pattern was used in order to correct for dead fibres within the field-of-view. Additionally, a blank sky exposure of 300s was obtained in between of the target exposure series. Arc lamp and continuum lamp exposures were acquired for each configuration directly after the target exposures for calibration purposes. Standard star exposures were taken for each night according to the instrumental setup.

The VIMOS instrument is a complex IFU with 1600 operating fibres in the high-resolution mode. These are split up into 4 bundles of 400 fibres densely projected onto each of the 4 spectrograph CCDs. Instrument flexure and cross-talk between adjacent fibres are important issues for VIMOS which need to be carefully taken into account in the data reduction. As this is not the case for the standard reduction pipeline provided by ESO, we used our own flexible reduction package R3D \citep{Sanchez:2006a} designed to reduce fibre-fed IFU raw data, complemented by custom Python scripts. The basic reduction steps performed with R3D include: Bias subtraction, visually checked fibre identification, fibre tracing, spectra extraction, wavelength calibration, fibre flatfielding and flux calibration.

Flexure of the instrument causes substantial shifts of the fibre traces in the cross-dispersion direction between the continuum lamp exposure and the science exposure. This is a severe problem for an accurate extraction of the science spectra. We measured the fibre trace positions in the individual science exposures and continuum lamp exposure by modelling the cross-dispersion profile at 5 different locations on the dispersion axis with multiple Gaussians. The cross-dispersion profiles were generated by co-adding 200\,pixels in the dispersion direction, preferentially encompassing bright sky lines, to increase the S/N. The offset between the science and continuum lamp fibre traces were measured to be in the range of $-2.5$ to 2.5 pixels. 2nd order Chebychev polynomials were used to extrapolate the offsets to the whole dispersion axis range.

In order to reduce the effect of cross-talk we used an iterative and fast algorithm that is part of the R3D package, based on Gaussian profile fitting to each of the fibres in the cross-dispersion direction. The centroids were fixed for each fibre to the position on the continuum lamp taking into account the position offsets due to the flexure. The width of the Gaussians was fixed. Details of the spectra extraction algorithm can be found in the R3D user guide\footnote{http://www.caha.es.sanchez/r3d/R3D\_user\_guide.pdf}. The width of the fibre profiles is different for each of the four CCDs as it depends on the optical calibration of the four independent VIMOS spectrograph units. Thus, we estimated the fibre dispersion separately for each unit by simultaneously fitting  each fibre with a Gaussian profile in the cross-dispersion direction at a single dispersion position for each CCD.

The continuum lamp exposure was used to create a fibreflat that corrects for the differences in the fibre-to-fibre transmission. Sky subtraction and the correction for the differential atmospheric refraction were done with custom Python scripts. Taking advantage of the dithered exposures we were able to increase the spatial sampling by a factor of 2 and to correct for dead fibres within the field-of-view. The flux calibrated datacubes were rescaled in flux to match with the QSO $V$-band photometry and corrected for Galactic extinction applying the \citet{Cardelli:1989} extinction curve with an extinction of $A_V=0.221$ \citep{Schlegel:1998} in the sightline to HE~1029$-$1401.

\section{IFU decomposition of host and QSO emission}
In order to study the emission from the stars and the gas in the host of HE~1029$-$1401 it is important to properly subtract the contribution of the bright QSO. Broad-band imaging studies of QSO hosts have successfully used a two-dimensional analytical modelling scheme to decompose the point-like nucleus and different spatial extended host components to study the properties of QSO host galaxies \citep[e.g.][]{McLure:1999,Kuhlbrodt:2004,Sanchez:2004b,Kim:2008,Jahnke:2009}. We have extended this method to model each monochromatic image of an IFU datacube in order to obtain the clean optical spectrum of the radio jet of 3C~120 \citep{Garcia-Lorenzo:2005,Sanchez:2004a,Sanchez:2006b} and to deblend the components of a gravitationally lensed quasar \citep{Wisotzki:2003}. 
\begin{figure}
 \resizebox{\hsize}{!}{\includegraphics[clip]{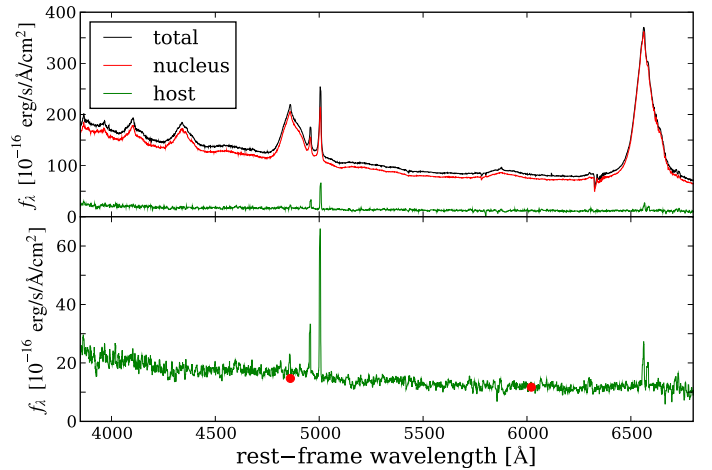}}
 \caption{Result of the decomposition process for HE~1029$-$1401. The integrated spectrum and the decomposed QSO and host spectra are shown in the top panel. A zoomed version on the host spectrum is provided in  the bottom panel. The red points indicate the host magnitudes in the V and R band measured by \citep{Jahnke:2004b}. They are offset by 0.5\,mag to put them on the same absolute scale as the IFU photometry. All spectra have been smoothed with a median filter of 7 pixels for display purposes.}
  \label{fig:decomp_spec}
\end{figure}

Broad-band imaging studies usually estimate the point spread function (PSF) empirically from stacked images of stars within the field-of-view. The field-of-view of current IFU instruments is generally too small so that no star can generally be captured simultaneously with the target. In light of time variable seeing, the QSO itself therefore needs to be used to construct a PSF. Fortunately, the spectral shape of the QSO is, in the absence of atmospheric dispersion,  exactly the same in each spatial pixel of the IFU, only scaled by a factor according to the PSF. In particular, the broad emission lines of type 1 QSOs are a unique feature of the point like nucleus, which can be used to empirically estimate a PSF for IFU data \citep{Jahnke:2004}. 

Having obtained a PSF for the IFU data from the broad emission lines and the morphological parameters of the host inferred from the HST images, we proceeded to model each monochromatic IFU narrow-band image by a linear combination of the PSF and host model. In this way we reconstructed a \textit{pure} QSO and host datacube containing the QSO and mean host spectrum. Afterwards, we subtracted the QSO datacube from the observed one to obtain a datacube uncontaminated by the QSO. The result of this decomposition process for HE~1029$-$1401 is presented in Fig.~\ref{fig:decomp_spec}.

We found the IFU spectrum of the QSO and host component to be of higher quality than the long-slit spectra presented by Let07. The stellar absorption lines around MgI\,$\lambda 5166$ are clearly visible in our host spectrum, and the noise in the spectrum is greatly reduced compared to the long-slit spectrum, due to the much larger galaxy area captured. IFU spectroscopy improves the quality of the decomposition due to the large spatial coverage of almost the entire host galaxy and the on-source PSF estimation based on QSO features. The decomposition of the long-slit spectra particularly suffered as the only available PSF star near HE~1029$-$1401 is 2\,mag fainter than the QSO (Let07).

\section{Analysis}\label{sect:data_analysis}
\subsection{The host spectrum and stellar velocity dispersion}\label{sect:host_spec}
\begin{figure}
 \resizebox{\hsize}{!}{\includegraphics[clip]{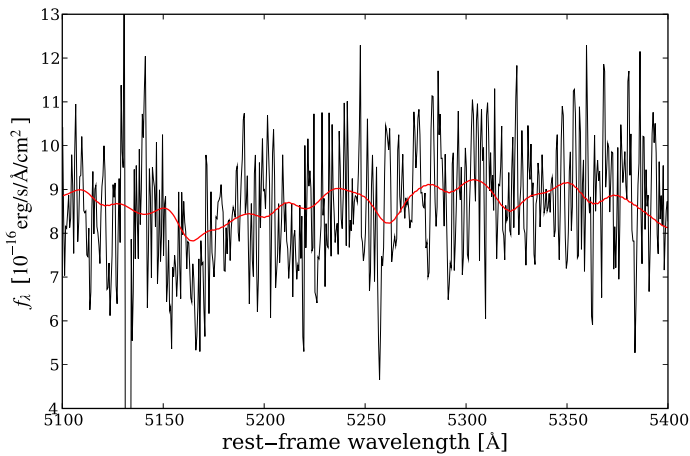}}
 \caption{Host spectrum of HE~1029$-$1401 in the rest-frame wavelength between 5100 and 5400\AA. The best-fitting model spectrum is overplotted as the red solid line. Note that the strong residual of the bright [\ion{O}{I}]\,5577 night sky emission line appearing at $\sim$5135\AA\ in the deredshifted spectrum was masked out for the modelling.}
  \label{fig:MgIb_fit}
\end{figure}
Comparing the $V$ and $R$ band apparent magnitudes reported by \citet{Jahnke:2004b} with the magnitudes inferred from our host spectrum, we found a constant offset of 0.5\,mag in both bands. This difference can be easily explained by the brightness variability of the QSO as we matched the overall IFU photometry to the archival magnitude of the QSO with a difference in observation time of several years. Note that our measured $V-R$ colour in the observed frame of 0.36\,mag is consistent with the $V-R$ broad-band colour of $0.4\pm0.05$\,mag measured by \citet{Jahnke:2004b}. 

While we found consistent colours with the previous broad-band study, we noticed a steep increase in the blue part of the total host spectrum.  This feature is even more pronounced for the host spectra within 1.5\arcsec\ around the nucleus. Whether this blue part of the spectrum originates from a young stellar population around the nucleus, or from scattered QSO light due to circumnuclear dust \citep[e.g][]{Antonucci:1985b,Zakamska:2005}, or whether it is an artifact of a systematic PSF mismatch in the decomposition process in that spectral region, we are currently unable to say. However, we do resolve the strong absorption lines around MgI\,$\lambda5166$ even at the low level of signal as shown in Fig.~\ref{fig:MgIb_fit}, so the stellar component has a significant contribution to the continuum emission.

We modelled the integrated host spectrum excluding the central 1.5\arcsec\ with a linear combination of high spectral resolution SSP spectra employing the STARLIGHT spectral synthesis code \citep{CidFernandes:2005,Mateus:2006}.
The SSP spectra with a spectral sampling of 0.3\AA\ were generated with the Sed@ code\footnote{Sed@ is a synthesis code included in the {\it Legacy Tool project} of the {\it Violent Star Formation Network}; see {\it Sed@ Reference Manual} at {\tt http://www.iaa.es/$\sim$mcs/sed@} for more information} as presented in  \citet{Gonzalez-Delgado:2005} with the following inputs: initial mass function from \citet{Salpeter:1955} in the mass range 0.1-120 $M_\odot$; the high resolution library from \citet{Martins:2005} based on atmosphere models from PHOENIX \citep{Hauschildt:1999,Allard:2001}, ATLAS9 \citep{Kurucz:1991} computed with SPECTRUM \citep{Gray:1994} and the ATLAS9 library computed with TLUSTY \citep{Lanz:2003}. Our input grid contains 39 SSP spectra with ages of 1,\,3,\,5,\,10,\,25,\,40,\,100,\,300,\,600,\,900,\,2\,000,\,5\,000 and \,10\,000\,Myr and metallicities of 0.004, 0.02 and 0.04\,Z.

Monte Carlo simulations were used to determine the errors of the best-fit parameters,  the flux-weighted stellar age $\langle t_*\rangle$ and metallicity $\langle Z\rangle$ at a normalisation wavelength of $\lambda_0=5075$\AA, the extinction ($A_V$) and the stellar velocity dispersion ($\sigma_*$). We generated 200 mock spectra of the host on the basis of the observed host spectrum combined with its wavelength dependent noise budget. These 200 artificial host spectra were analysed with STARLIGHT exactly as the observed host spectrum itself. Finally, a linear combination of only 5--7 SSP spectra was sufficient for modelling all spectra. The other SSP base spectra of the grid were discarded from the modelling in the first iteration by STARLIGHT as their total contribution to the flux density at the normalisation wavelength were less then $2$\%. The resulting parameters were $\langle t_* \rangle =5\pm0.5$\,Gyr, $\langle Z \rangle = 0.024\pm0.003$, $A_V = 0.5\pm0.2$ and $\sigma_* = 320\pm 90\,\mathrm{km/s}$ (corrected for the instrumental resolution of $\sim$42\,km/s). We repeated the same analysis excluding the blue part of the host spectrum (3800--4700\AA\ rest-frame wavelength) and arrived at consistent results within the errors. To our knowledge this is the first time that $\sigma_*$ was estimated for the host galaxy of such a luminous QSO. \citet{Jahnke:2004b} derived a younger stellar population with 0.7-2\,Gyr based on SSP modeling of the $VRIJHK$ multi-colour spectral energy distribution. A direct comparison with our result needs to be taken with caution due to the completely different analysis method used to infer the characteristic age of the stellar population. However, we note that the distribution of stellar populations that contribute to the composite spectrum is clearly bimodal and consists of a young stellar population with ages $<$100\,Myr ($53\pm5\%$ of the total flux at $\lambda_0$) in addition to an old stellar population with 10\,Gyr age ($47\pm5\%$ of the total flux at $\lambda_0$). This indicates the presence of a young stellar component if the main contribution to the blue continuum of the spectrum is of stellar origin and not due to a featureless continuum that can hardly be distinguished from such a young stellar component \citep[e.g][]{Storchi-Bergmann:2000}.

\subsection{The QSO spectrum}
\begin{figure}
\resizebox{\hsize}{!}{\includegraphics[clip]{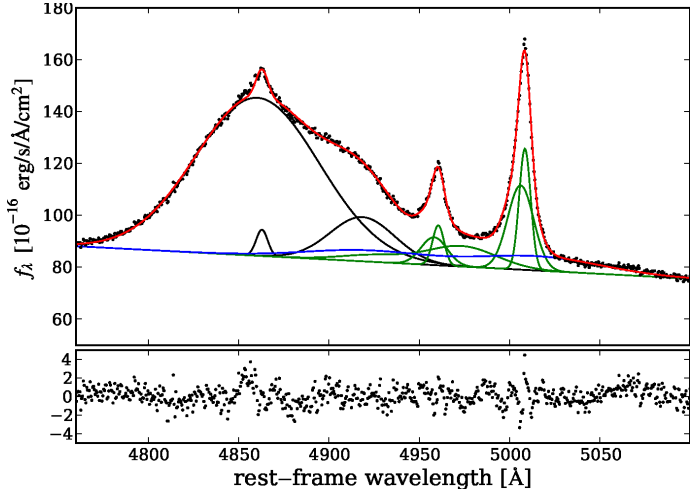}}

\caption{Multi-component fit to the H$\beta$-\Ox\ spectral region of the QSO spectrum. The best-fit model is indicated by the red solid line and the individual Gaussian components are plotted above the local linear continuum with the following colour coding: H$\beta$ line (black), \Ox\ doublet lines (green), \Fe\ doublet lines (blue). The residuals of the model are shown below with a refined scaling. }
\label{fig:qso_spectrum}
\end{figure}
\begin{figure}
\resizebox{\hsize}{!}{\includegraphics[clip]{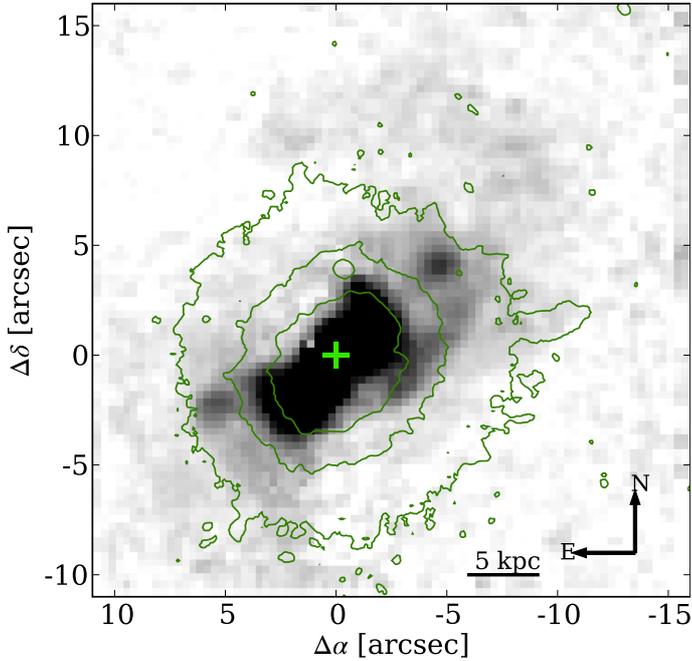}}
\caption{Nucleus-subtracted continuum-free \Ox\,$\lambda 5007$ narrow-band image. The 20\AA\ wide band was centred on the nuclear \Ox\,$\lambda 5007$ wavelength. A green cross marks the position of the QSO. The contours indicate the position and morphology of the host in comparison to the ionised gas morphology.}
\label{fig:O3_image}
\end{figure}
\begin{figure}
\centering
\includegraphics[height=16cm,clip]{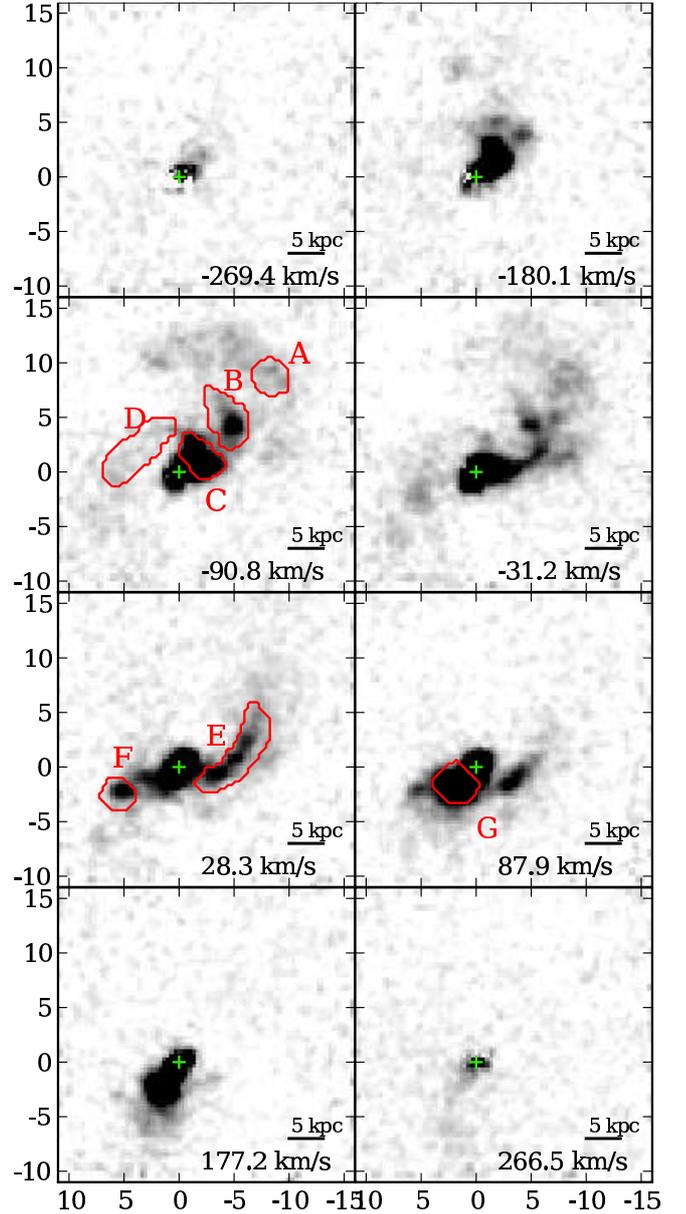}
\caption{\Ox\ channel maps created from monochromatic datacube slices corresponding to rest-frame radial velocities of $-269.4$, $-180.1$, $-90.8$, $-31.2$, $+38.3$, $+87.9$, $+177.2$, $+266.5$\,km/s, respectively. The main distinct emission line regions are labelled alphabetically from A to G and the red boxes indicate regions that were co-added to obtain their characteristic spectrum (Fig.~\ref{fig:region_specs}). The position of the QSO is marked in each map by the green cross.}
\label{fig:O3_channel}
\end{figure}
The top panel of Fig.~\ref{fig:decomp_spec} shows the QSO spectrum of HE~1029$-$1401 in the full wavelength range covered by our observations. We now focus on the analysis of the \Hb-\Ox\ region using a multi-component fitting scheme to deblend the various emission lines in this spectral region. Our model consists of several Gaussian components and a simple linear relation to approximate the local continuum. The best-fit model to the spectrum and its individual components are shown in Fig.~\ref{fig:qso_spectrum}.

 The \Hb\ line exhibits an asymmetric line profile with enhanced emission on its red wing, which can be well described by a redshifted Gaussian component. From the model we measured a line width of the broad \Hb\ line of 6195$\pm$80\,km/s (FWHM) and 2352$\pm$30\,km/s ($\sigma$). Furthermore, the \Ox\ line profile is asymmetric to the blue side requiring a second Gaussian component. The two \Ox\ components are separated by 120$\pm$50 km/s in rest-frame and have FWHM values 421$\pm$50\,km/s and 975$\pm$129\,km/s, respectively.  Additionally, a third extremely broad and blueshifted component is required to reach an acceptable fit. We are not sure whether this component is physical, representing a true extension of the asymmetric \Ox\ line profile, or whether  it should be rather attributed to the complex and asymmetric line profile of the H$\beta$ line. The continuum flux density at 5100\AA\ (rest-frame) is  $(93.4\pm 11.6) \times 10^{-16}\,\mathrm{erg}\mathrm{s}^{-1}\mathrm{cm}^{-2}\mathrm{\AA}^{-1}$  corresponding to a luminosity of $\log(\lambda L_{\lambda}/[\mathrm{erg/s}]) = 44.97\pm0.05$.

With these measurements we estimated a black-hole mass from the single epoch QSO spectrum using the virial method. This calculation is based on the empirical relation between the broad-line region radius and the continuum luminosity at 5100\,\AA\ calibrated by \citet{Bentz:2006}, and the prescription by \citet{Collin:2006} to infer the velocity of the broad line clouds from the H$\beta$ line dispersion adopting a scale factor of $3.85$. This yields a black hole mass of $\log(M_\mathrm{BH}/M_{\sun})= 8.7$ for HE~1029$-$1401. The measurement errors for the black hole mass are much smaller in this case than the systematic errors of the method, thus we adopted a canonical error of 0.3\,dex. 
Assuming the $M_\mathrm{BH}$-$\sigma_*$ relation of \citet{Tremaine:2002} our measured stellar velocity dispersion $\sigma_*$ predicts a black hole mass of $\log(M_\mathrm{BH}/M_{\sun})=8.9_{-0.4}^{+1.0}$, which is perfectly consistent with our virial black hole mass estimate.
\begin{figure*}
\centering
\includegraphics[height=24cm,clip]{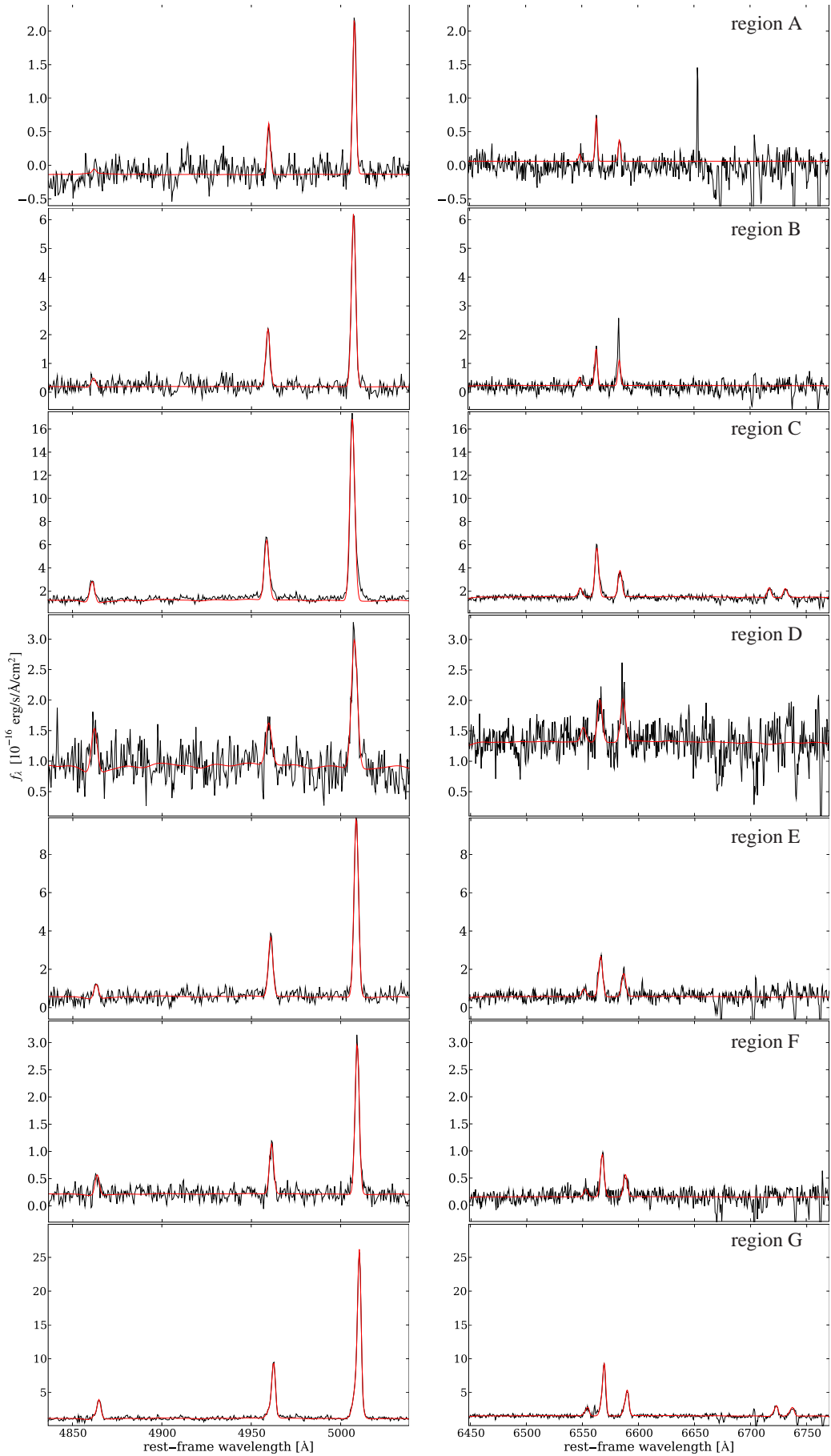}
\caption{Co-added spectra of the regions A-G (as defined in Fig.~\ref{fig:O3_channel}) are shown from top to bottom split up into two wavelength ranges bracketing \Hb\ (left panels) and \Ha\ (right panels). The black line corresponds to the observed spectrum and the red line to the fitted model. Two systems of Gaussians were required to model the spectrum of region G. The apparent mismatch between the model and the \Ni\,$\lambda6585$ line in region B is due to an artifact, but the model is still constrained due to the doublet nature and the kinematical coupling of the lines.}
\label{fig:region_specs}
\end{figure*}

Adopting a bolometric correction of $L_\mathrm{Bol} = 9\times \lambda L_\lambda(5100\AA)$ \citep{Kaspi:2000}, we estimated a bolometric luminosity of $\log(L_\mathrm{Bol}/[\mathrm{erg/s}]) = 46.0\pm0.2$  and an Eddington ratio of $\log(L_\mathrm{Bol}/L_\mathrm{Edd}) = -0.9\pm0.2$ for the QSO.

\subsection{Ionised gas distribution and channel maps}\label{sect:O3_map}
In Fig.~\ref{fig:O3_image} we present an \Ox\,$\lambda 5007$ (hereafter \Ox) narrow-band image ($20\AA$ bandwidth) extracted from the datacube after subtracting the QSO contribution. Continuum emission from the host galaxy was previously subtracted to obtain a pure emission-line image. This is the first two-dimensional map of the ionised gas distribution around HE~1029$-$1401.

The image reveals ionised gas on scales of several kpcs. We find a highly structured distribution of the ionised gas. The brightest structure is a symmetric circumnuclear bicone suggestive of an ionisation cone by the QSO. Furthermore, we detect ionised gas in several knots and arms and in a faint giant arc-like feature north-west to the galaxy centre at a projected distance of 16\,kpc. Most of the distinct emission-line regions are well located inside the visible light distribution of the early-type host galaxy when compared with the HST image in Fig.~\ref{fig:HST_image}. The prominent bicone-like structure is roughly aligned with the major axis of the host galaxy, although the orientation of the latter is not well constrained due to the low ellipticity of the host. 

The integrated extended \Ox\ line flux is $(328\pm88)\times 10^{-16}\,\mathrm{erg}\,\mathrm{s}^{-1}\mathrm{cm}^{-2}$ corresponding to an \Ox\ luminosity of $\log (L_{\Ox}/[\mathrm{erg/s}]) = 41.78\pm0.12$. This is one of the most luminous extended emission-line regions (EELRs) around a radio-quiet QSO compared to other radio-quiet, e.g. in the sample of \citet{Stockton:1987}.

In  Fig.~\ref{fig:O3_channel} we present \Ox\ channel maps corresponding to monochromatic images at specific radial velocities with respect to the rest-frame of the objects. In this way we can identify low-surface brightness emission-line regions that are relatively faint in the integrated line image or that spatially overlap but are separated in velocity space. The main emission line regions are labelled alphabetically from A to G and their characteristic spectra were obtained by co-adding several spatial pixels within the boundary indicated by the red boxes. These regions are a symmetric bicone-like structure around the nucleus (regions C and G), two arm-like structures (regions D and E), two emission-line knots (regions B and F) and part of the giant arc-like feature $\sim$16\,kpc north-west of the nucleus (region A). We excluded the central $\sim$1\arcsec\ region around the QSO from these spectra because of  significant residuals of the QSO subtraction.

\subsection{Spectral analysis of specific regions}\label{subsect:spec_regions}
Co-added spectra of the regions A--G as defined in the previous subsection are shown in Fig.~\ref{fig:region_specs}. These were split up into two narrow wavelength ranges bracketing the \Hb\ and \Ha\ emission lines. No signatures of the broad emission lines of the QSO are visible in any of the extracted spectra after the spectral decomposition. The \Hb, \Ox\,$\lambda\lambda4959,5007$, \Ha\ and \Ni\,$\lambda\lambda6549,6585$ emission lines are clearly detected in most of the selected regions. The \Su\ $\lambda\lambda6716,6731$ emission lines are only visible in the spectra of region C and G.

  We modelled each spectrum with Gaussian profiles for the emission lines and scaled the best-fitting stellar population spectrum (cf. Sect.~\ref{sect:host_spec}) to the continuum. We found that stellar absorption lines have no significant effect on the Balmer emission line. The flux ratio of the \Ox\ and \Ni\ doublets were fixed to their theoretical values and all emission lines were kinematically coupled to have the same redshift and line dispersion. We utilised Monte Carlo simulations to estimate realistic errors for all free fit parameters. 100 mock spectra were generated for each spectrum according to the noise of the spectrum as estimated from the adjacent continuum. These mock spectra were consistently analysed and the standard deviation of the resulting distribution for each fitted parameter was taken as its $1\sigma$ error. A $3\sigma$ upper limit is given for the flux of the H$\beta$ line in region A as it falls below the detection limit. The measured  emission line fluxes are given in Table~\ref{tab:lines}.

\begin{table*}
\centering
\begin{footnotesize}
\caption{Emission line fluxes for the H$\beta$, \Ox\,$\lambda5007$, H$\alpha$,\Ni\,$\lambda 6583$ and the \Su\ $\lambda\lambda6716,6731$ lines as measured from Gaussian fits to the spectra of the region A to G.}
\label{tab:lines}
\begin{tabular}{ccccccc}\hline\hline \noalign{\smallskip}
Region & H$\beta$ & \Ox$\,\lambda5007$ & H$\alpha$ & \Ni$\,\lambda 6583$ & [\ion{S}{II}]\,$\lambda 6716$ & [\ion{S}{II}]\,$\lambda 6731$ \\
& \multicolumn{6}{c}{line fluxes in units of $10^{-16}\,\mathrm{erg}\,\mathrm{s}^{-1}\,\mathrm{cm}^{-2}$} \\ \noalign{\smallskip} \hline \noalign{\smallskip}
A & $<0.5$ & $5.9\pm 0.2$ & $1.9\pm 0.2$ & $1.1\pm 0.1$ & -- & -- \\
B & $1.0\pm 0.2$ & $19.5\pm 0.3$ & $4.2\pm 0.3$ & $4.7\pm 0.2$ & -- & -- \\
C & $7.2\pm 0.2$ & $56.8\pm 0.4$ & $21.7\pm 0.3$ & $11.9\pm 0.3$ & $4.7\pm 0.3$ & $ 3.8\pm 0.3$ \\
D & $2.9\pm 0.4$ & $8.5\pm 0.5$ & $4.9\pm 0.5$ & $4.2\pm 0.5$ & -- & -- \\
E & $2.7\pm 0.4$ & $32.0\pm 0.5$ & $10.5\pm 0.4$ & $6.1\pm 0.3$ & -- & -- \\
F & $1.3\pm 0.1$ & $9.0\pm 0.1$ & $3.8\pm 0.2$ & $2.0\pm 0.2$ & -- & -- \\
G (narrow) & $5.7\pm 0.5$ & $52.7\pm 1.7$ & $19.1\pm 3.4$ & $7.5\pm 2.3$ & $2.8\pm 1.1$ & $ 1.9\pm 0.9$ \\
G (broad) & $5.2\pm 0.6$ & $29.9\pm 1.8$ & $14.7\pm 2.5$ & $10.6\pm 2.2$ & $4.8\pm 1.2$ & $ 4.5\pm 1.1$  \\ \noalign{\smallskip} \hline
\end{tabular}

\end{footnotesize}
\end{table*}

We noticed that all emission lines in the spectrum of  region G display an asymmetric line profile. A two component Gaussian model for each line provided an excellent fit to the data. The appearance of a second system of kinematically distinct emission-line in region G will be further analysed later in Sect.~\ref{subsect:sigma_gas}.

\section{Results}\label{sect:results}
\subsection{Source of ionisation and gas metallicity}\label{subsect:ion}
\begin{figure}
\resizebox{\hsize}{!}{\includegraphics[clip]{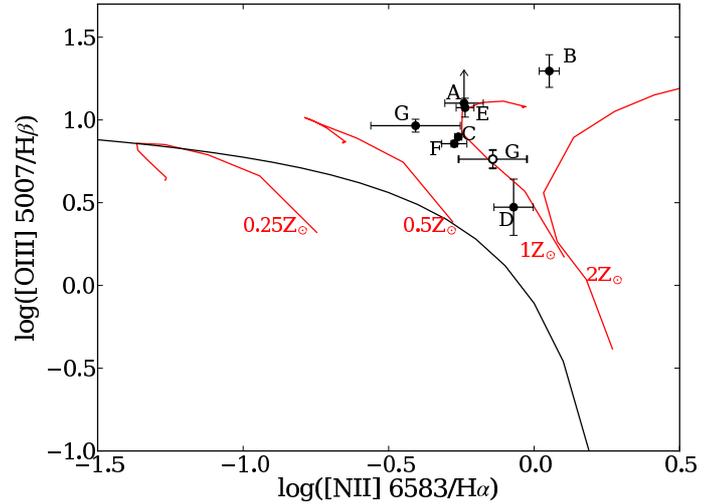}
}
\caption{Diagnostic diagram for the emission-line ratios of the regions A-G in HE~1029$-$1401. The narrow emission-line ratios are indicated by black circles and the broad component in region G is indicated by the black opened circle.  The commonly used demarcation curve between AGN and \ion{H}{ii} regions by \citet{Kewley:2001} is indicated as the black solid line. Dusty radiation pressure-dominated photoionisation models \citep{Groves:2004} are indicated by the red tracks for four gas-phase metallicities with $0.25Z_{\sun}$, $0.5Z_{\sun}$, $1Z_{\sun}$ and $2Z_{\sun}$ with physical parameters as described in the text.}
\label{fig:diagnostic_diagram}
\end{figure}
To infer the dominant ionisation source for the gas we employ  the commonly used emission-line ratio diagnostic diagram \Ox\ $\lambda5007$/$H\beta$ vs. \Ni $\lambda6582$/H$\alpha$ \citep{Baldwin:1981,Veilleux:1987}. The diagram distinguishes gas being ionised by the hard UV radiation field of an AGN from gas ionised by radiation from hot stars in star forming \ion{H}{II} regions. A theoretically motivated and conservative boundary between both mechanisms was derived by \citet{Kewley:2001}. The diagnostic diagram is shown in Fig.~\ref{fig:diagnostic_diagram} for the emission-line ratios inferred from the spectra of region A--G.

We found that photoionisation by star forming regions can be excluded for all regions as the inferred line ratios are above from the Kewley et al. demarcation curve. Of course, this does not imply there is no current star formation in the host, only that AGN ionisation of the ISM dominates strongly over any other ionisation mechanisms. We can also exclude shock ionisation based on the kinematics and the emission-line ratios. The line ratios for the broad shocked gas components in powerful radio galaxies are observed to have a much lower ionisation state with \Ox/\Hb$\sim$2--4 \citep[e.g.][]{Villar-Martin:1999} than observed here. Furthermore,  shock-ionisation models \citep{Dopita:1995,Allen:2008} indicate that an \Ox/\Hb\ line ratio around 10 would require shock velocities $>500$\,km/s for which we find no evidence in our observations (cf. Sect.~\ref{subsect:kin}). Thus, photoionisation by the central QSO is the dominant ionisation mechanism for the gas in HE~1029$-$1401 \textit{at least} within a radius of 16\,kpc around the nucleus.

 In order to infer the metallicity of the gas we compared the emission-line ratios with the dusty radiation pressure-dominated photoionisation models by \citet{Groves:2004}. These models are represented by red tracks in Fig.~\ref{fig:diagnostic_diagram} corresponding to gas-phase metallicities of $0.25Z_{\sun}$, $0.5Z_{\sun}$, $1Z_{\sun}$ and $2Z_{\sun}$ assuming an electron density of $n=100\mathrm{cm}^{-3}$ for the emitting clouds and a power-law index $\alpha=-1.4$ for the ionising QSO continuum. The line ratios do not strongly depend on the electron density, so we chose the model grid with the lowest density close to the estimated value as discussed in the next section. Note that the true metallicities of the photoionisation models are a factor of 2 higher than the gas-phase metallicities, since half of the overall metal content is depleted into dust in the models. The comparison with the photoionisation models revealed that the emission-line ratios follow quite closely the model with a solar metallicity. Only the gas in region B appears to  have a systematically different ionisation state and/or metallicity. 
\begin{figure*}
\sidecaption
 \includegraphics[width=5.5cm,height=4.5cm,clip]{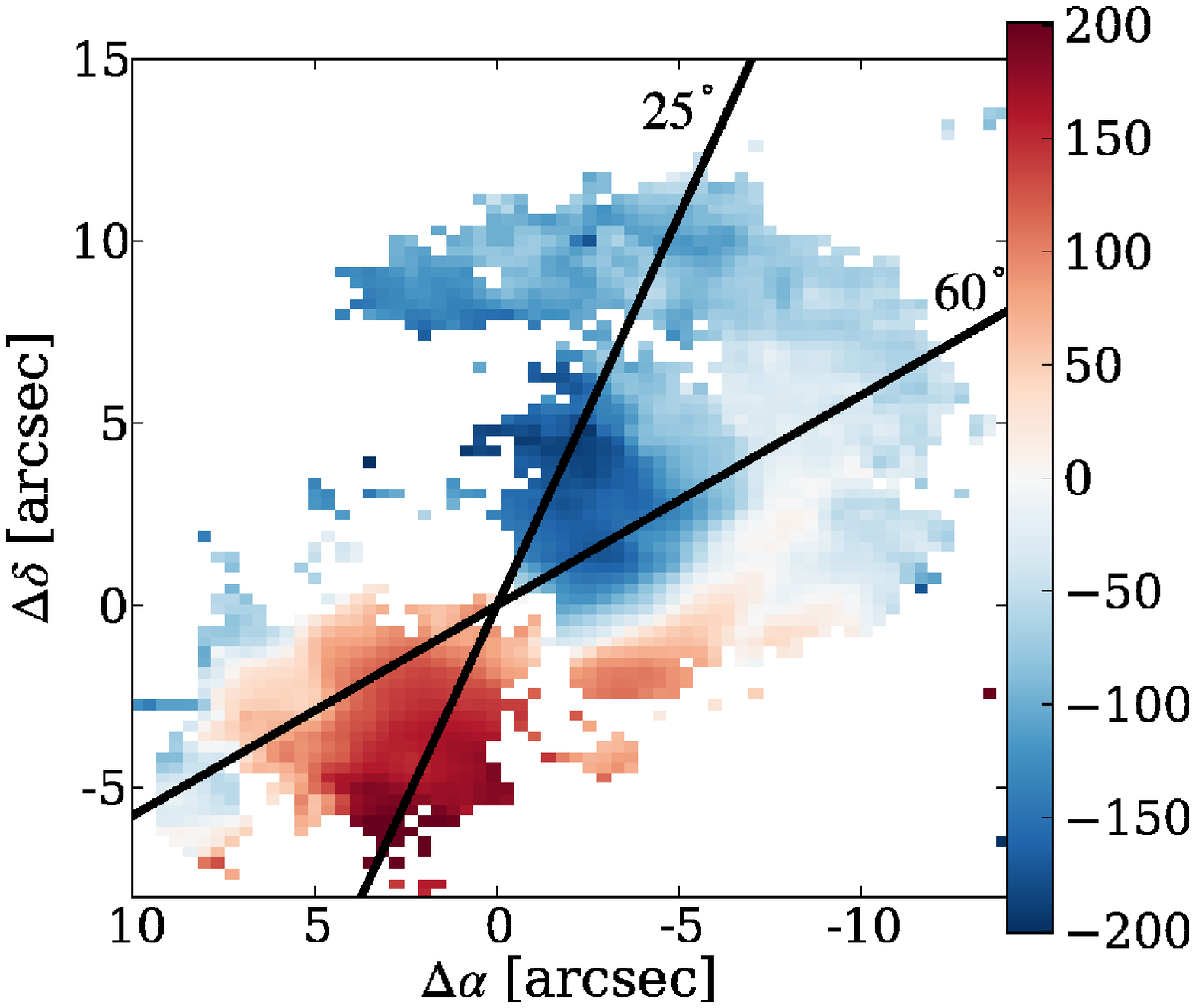}
\includegraphics[width=6.5cm,height=4.5cm,clip]{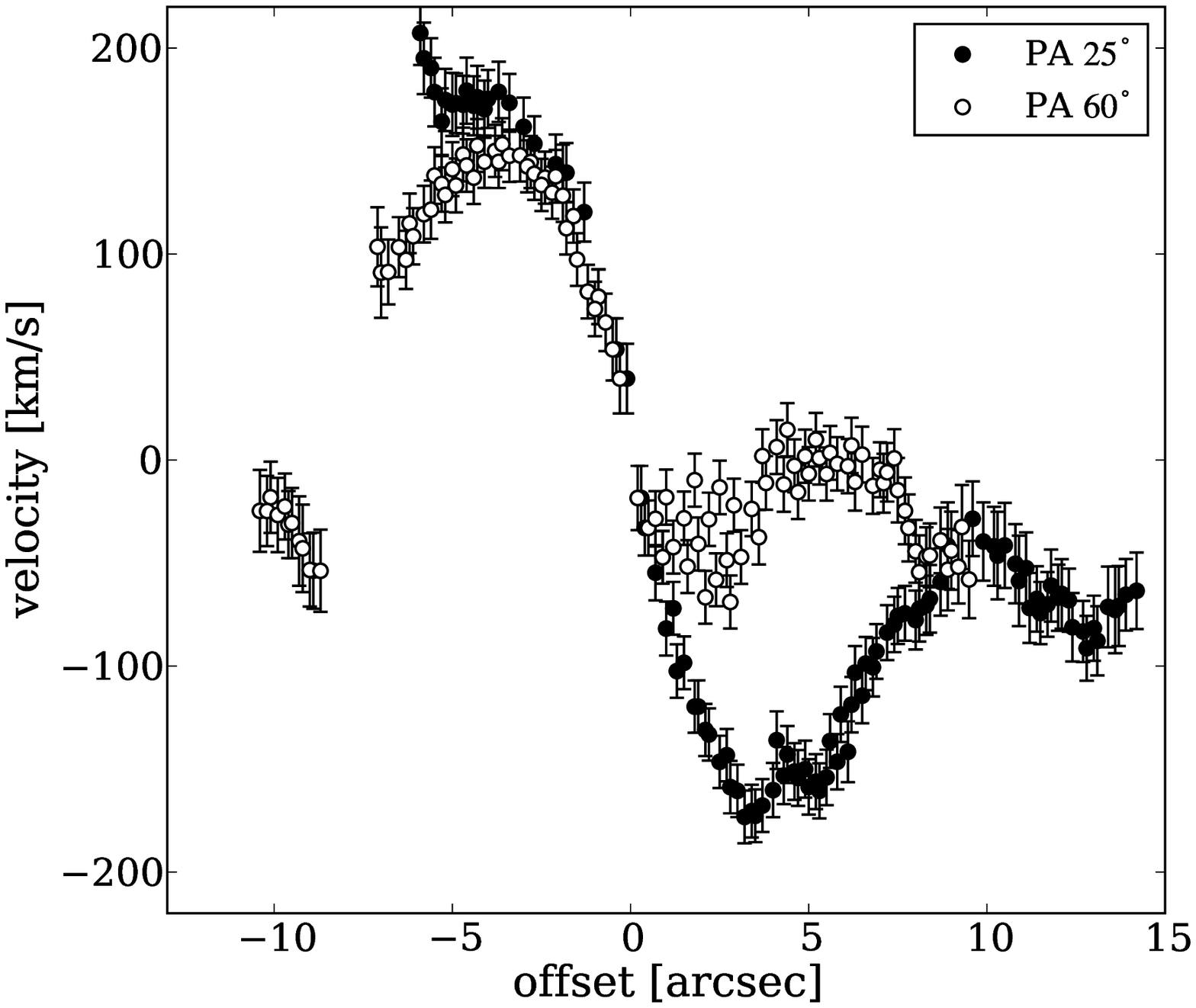}
\caption{\emph{Left panel:} \Ox\ velocity map of HE~1029$-$1401 with respect to its estimated systemic velocity ($25670$\,km/s). The two black lines correspond to position angles of $25^\circ$ and $60^\circ$, respectively. \emph{Right panel:} Velocity curves extracted for these two orientations from the maps. Negative offsets correspond to the South-East side and positive offsets to the North-West side of the nucleus.}
\label{fig:O3_vel}
\end{figure*}
\begin{figure*}
 \sidecaption
\includegraphics[width=5.5cm,height=4.5cm,clip]{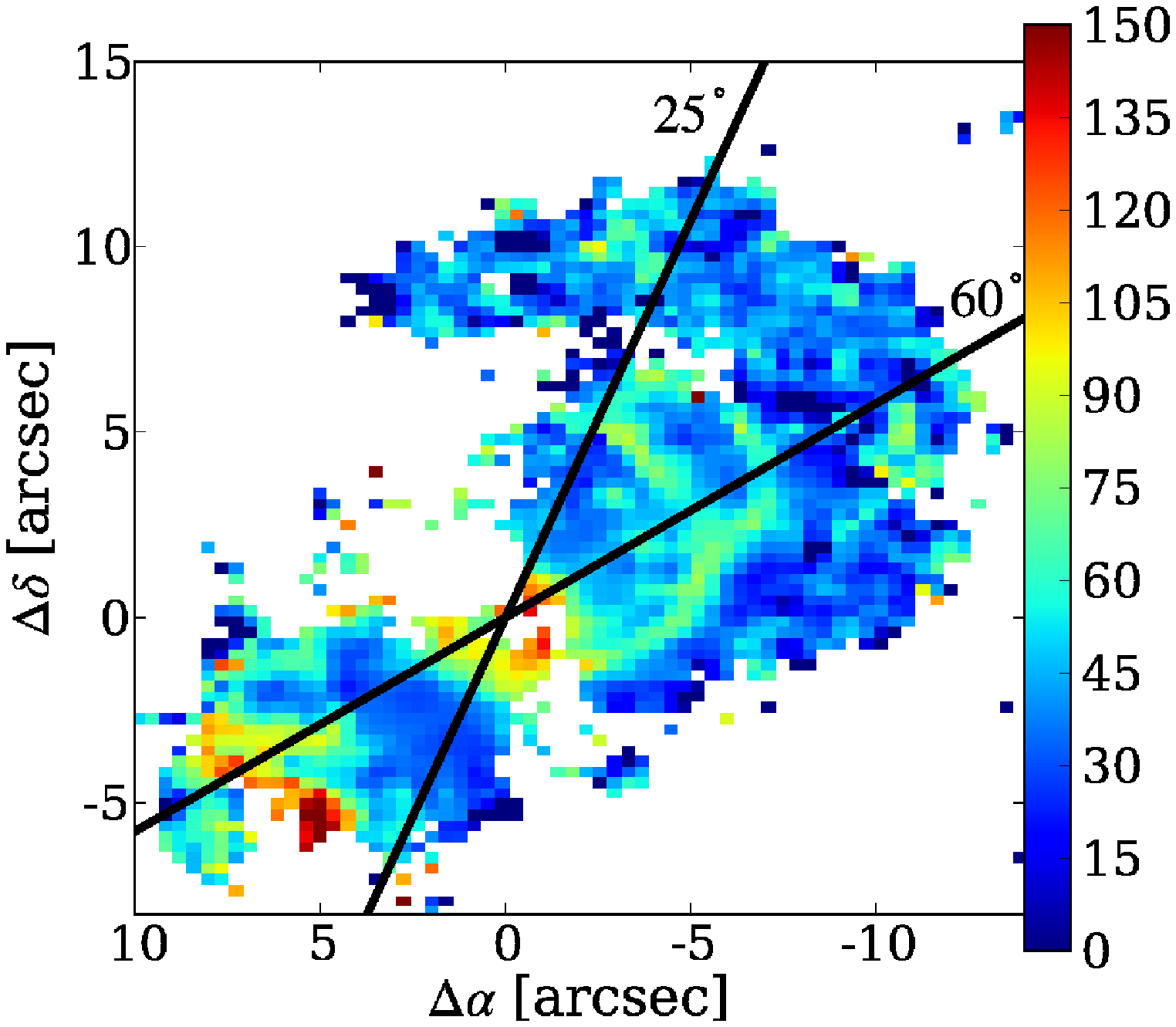}
\includegraphics[width=6.5cm,height=4.5cm,clip]{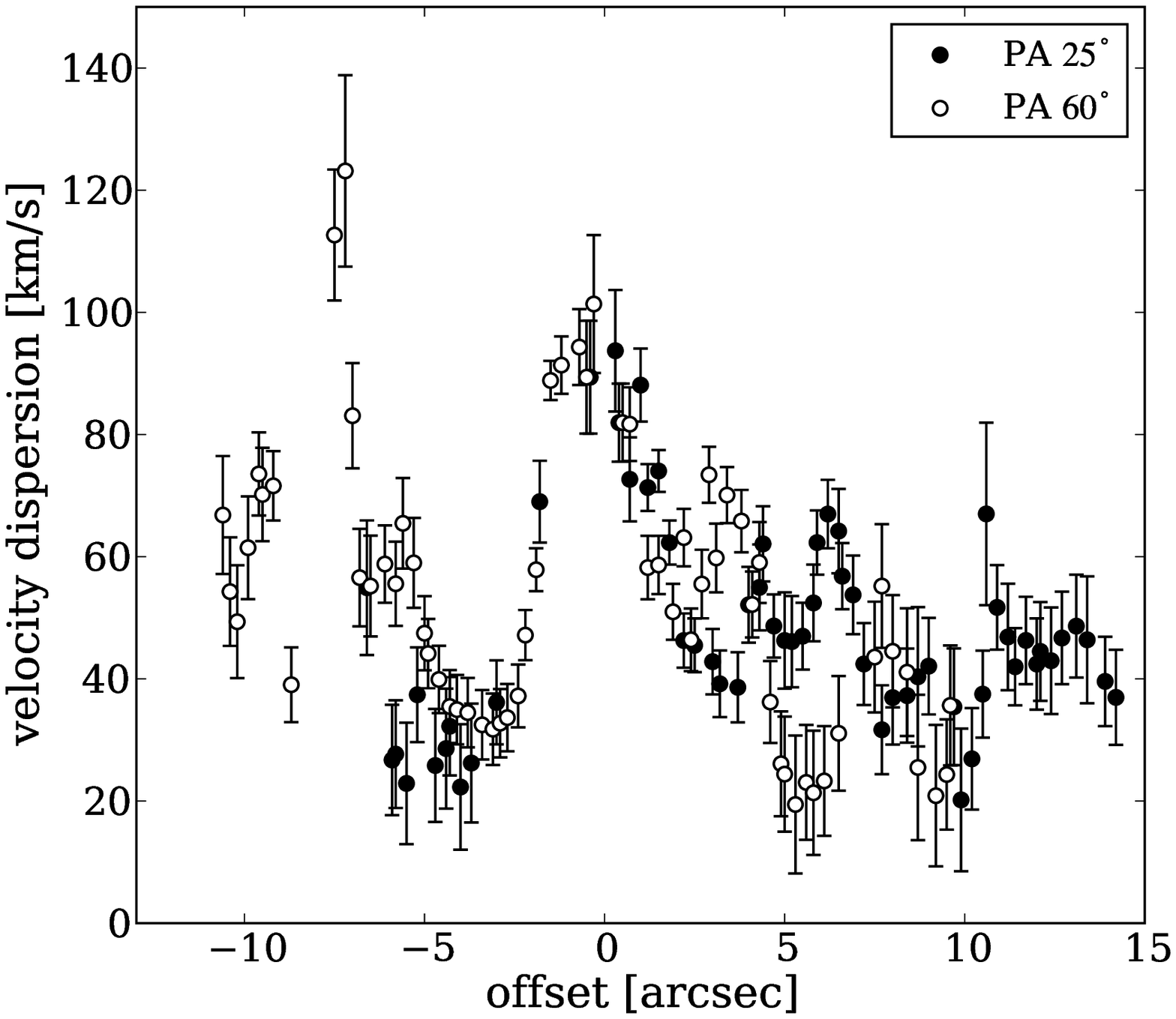}
\caption{Same as Fig.~\ref{fig:O3_vel}, but for the \Ox\ velocity dispersion, corrected for the spectral resolution of the instrument.}
\label{fig:O3_disp}
\end{figure*}
\subsection{Total ionised gas mass}\label{subsect:gas_mass}
To estimate the total ionised gas mass in HE~1029$-$1401 we followed the prescription and assumptions of \citet{Osterbrock:2006}. By measuring the luminosity of the H$\beta$ recombination line one can estimate the ionised gas mass $M_\mathrm{ion}$ using the following formula:
\begin{equation}
 M_{\mathrm{ion}} = \frac{1.4 m_\mathrm{p}}{n_e\alpha_{\mathrm{H}\beta}^{\mathrm{eff}} h\nu_{\mathrm{H\beta}}}L_{\mathrm{H}\beta}\ ,\label{eq:ion_mass}
\end{equation}
where $m_\mathrm{p}$ is the proton mass, $h$ is Planck's constant, $n_e$ is the electron density, $\alpha_{\mathrm{H}\beta}^{\mathrm{eff}}$ is the effective recombination coefficient for H$\beta$  and $L_{\mathrm{H}\beta}$ is the $H\beta$ line luminosity. Assuming Case B recombination, the low-density limit and a typical gas temperature of $T\approx 10,000$\,K for ionised nebulae, we end up with

\begin{equation}
 M_{\mathrm{ion}} \approx 10^{7}\,\left(\frac{100\,\mathrm{cm}^{-3}}{n_\mathrm{e}}\right)\left(\frac{L_{\mathrm{H}\beta}}{10^{41}\,\mathrm{erg}\,\mathrm{s}^{-1}}\right)
\end{equation}

In our case  we were able to measure the electron density sensitive \Su\,$\lambda6716/\lambda6731$ emission-line ratio only in high-surface brightness regions of the bicone structure (region C and G). The line ratios in region G are rather uncertain due to the line blending of a narrow and broad component, so that we estimated the electron density from region C only. We measured a \Su\,$\lambda6716/\lambda6731$ line ratio of  $1.2\pm0.1$  for region C. This ratio implies an electron densities  of $200_{-110}^{+150}\mathrm{cm}^{-3}$  \citep[][p. 123]{Osterbrock:2006} assuming an electron temperature of 10\,000\,K.

 Since the narrow H$\beta$ line is the weakest in all of the spectra, we used the \Ox\ emission line as a surrogate for the \Hb\ emission assuming a line ratio of $\Ox/\Hb\sim10$ (cf. Fig.~\ref{fig:diagnostic_diagram}). With the integrated $\Ox$ line luminosity measured from the \Ox\ narrow-band image (c.f. Sect.~\ref{sect:O3_map}) and our adopted electron density, we estimated an ionised gas mass of $M_\mathrm{ion}=3_{-2}^{+6}\times10^6\,M_{\sun}$.  Note that we did not take reddening due to dust in the host galaxy into account, which would increase $L_{\mathrm{H}\beta}$ and hence the ionised gas mass. Our inferred ionised gas mass should be taken as a rough estimate due to the various assumptions needed and can only be used at an order-of-magnitude level.

\subsection{Global gas kinematics}\label{subsect:kin}

In order to obtain full 2-D velocity and velocity dispersion maps of the ionised gas we fitted the \Ox\ doublet lines as they are the brightest emission lines in all of the individual 6400 spectra of the nucleus-subtracted datacube. The measured emission-line centroids were converted into radial velocities with respect to the systemic redshift for each spatial pixel to construct a 2-D velocity map (Fig.~\ref{fig:O3_vel} left panel). A different visualisation of the velocity field are `long-slit' velocity curves (Fig.~\ref{fig:O3_vel} right panel) extracted  from our velocity map for the position angles (PAs) $25^\circ$ and $60^\circ$. We chose these two PAs such that one shows the velocity curve of the bi-cone structure and the other to cross the arm-like feature. Similarly, we present a 2-D \Ox\ velocity dispersion map (Fig.~\ref{fig:O3_disp} left panel), corrected for the spectral resolution of the instrument, and the corresponding `long-slit' curves (Fig.~\ref{fig:O3_disp} right panel). The gaps in the velocity and dispersion curves around the circumnuclear region are due to the strong residuals of the decomposition process which prevent a reliable measurement of the extended emission lines. 

The 2-D kinematic information from our IFU data draws a much clearer picture than previous long-slit studies of this object.  While the \Ox\ velocity curve at a PA of $25^\circ$ appears like a symmetric rotation curve up to 5\arcsec\ from the galaxy centre, it is  highly disturbed at larger distances. At a PA of $60^\circ$ the velocity curve crosses the arm-like structure detected in the \Ox\ flux distribution which clearly breaks the symmetry seen at a PA of $25^\circ$. We measured a maximum amplitude of  $\pm 170$\,km/s for the radial velocities at a distance of 4.8\,kpc  ($3\arcsec$) from the nucleus, which is consistent with the measurements by Let07 and Jah07 based on long-slit spectroscopy only. However, the long-slit velocity curve presented by Let07 covers only $\pm 4\arcsec$ which is less than half of the region we studied with our IFU data. Therefore, they miss the turnover  of the velocity curve at 4\arcsec--6\arcsec from the nucleus. Another major kinematic component is the prominent distortion in the rotational velocity field on the West side of the nucleus, which was missed by the long-slit data of Let07 and Jah07.

The measured velocity dispersion ($\sigma$) of the gas after correcting for the instrumental resolution is $\sim30$\,km/s on the red- and $\sim$40\,km/s on the blue-shifted side of the velocity curve. On both sides the  $v/\sigma$ ratio is significantly greater than 1, which indicates rotational support of the gas in a disc. This is consistent with the conclusion of Jah07. The velocity dispersion increases significantly up to $\sim80$\,km/s between the region of the gas disc and the major kinematic distortion on the West side. This is most probably related to the superposition of the two kinematic components as we find asymmetric and double-peaked emission-lines in this particular region. In the circumnuclear region the velocity dispersion rises steeply towards the centre. We will  study the kinematics of this region in greater detail in the next section.
\begin{figure}
\resizebox{\hsize}{!}{\includegraphics[clip]{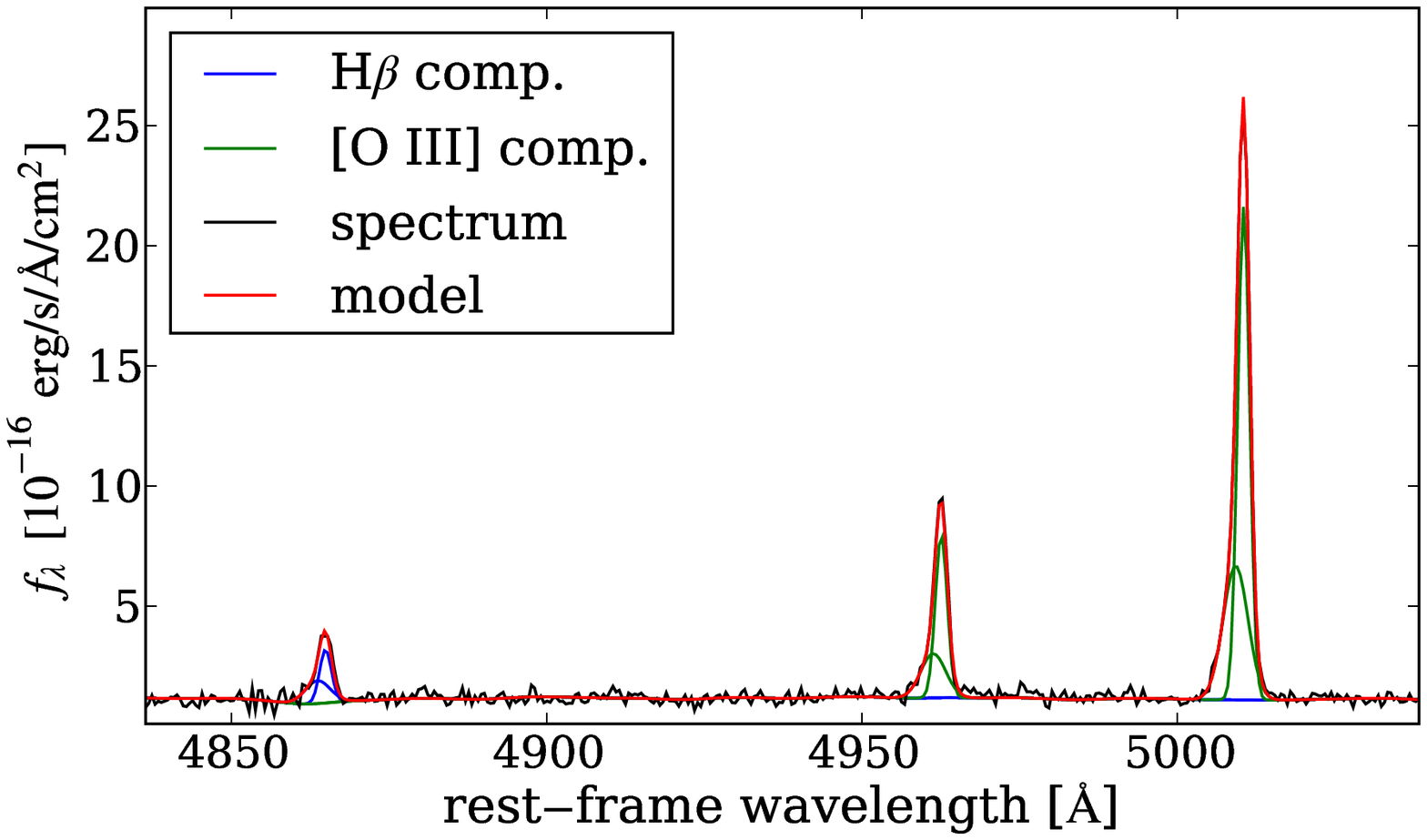}}
\resizebox{\hsize}{!}{\includegraphics[clip]{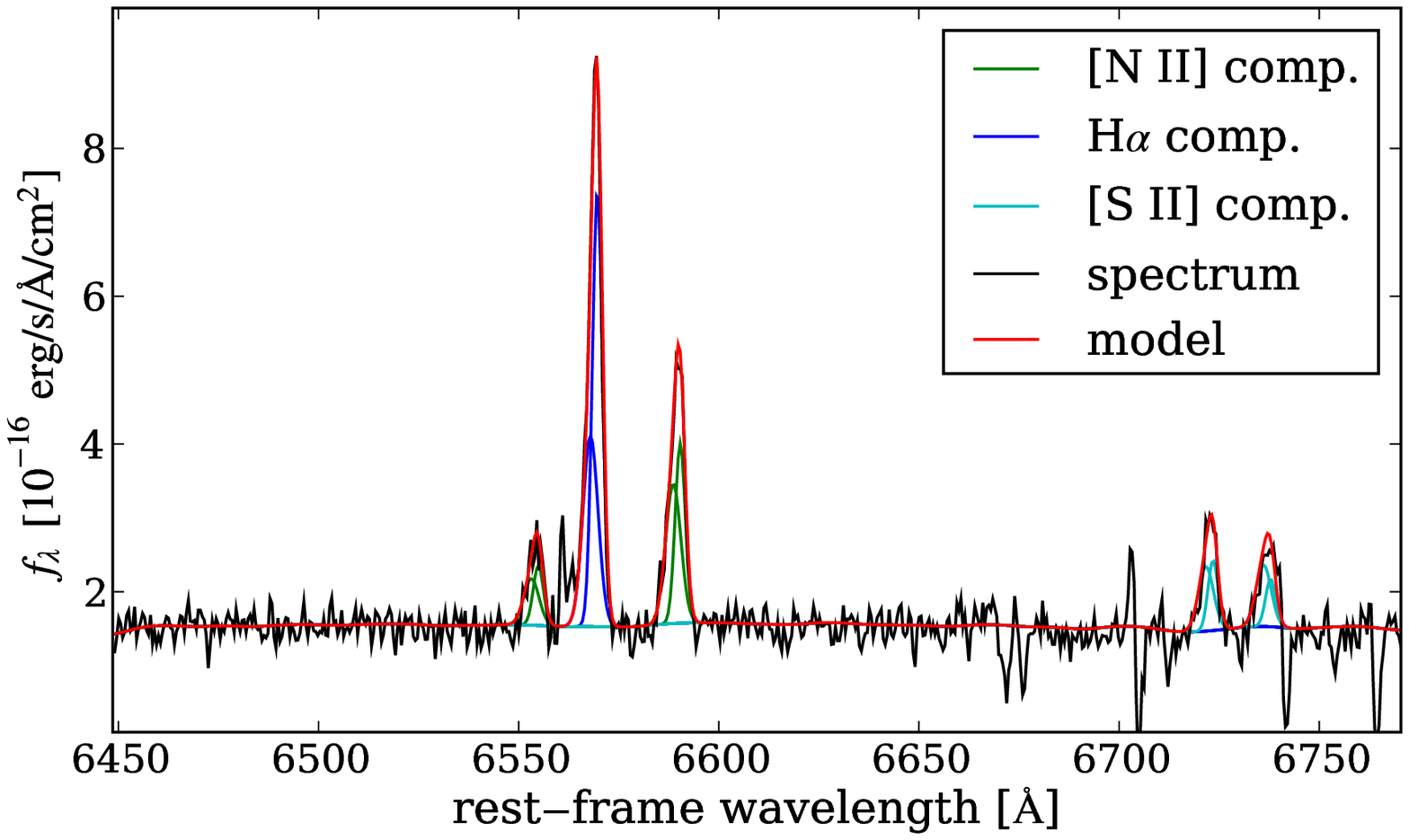}}
\caption{Detailed modelling of the emission lines in the spectrum of region G. The emission lines around H$\beta$ are shown in the top panel and the ones around H$\alpha$ in the bottom panel. The spectrum and its model together with the individual Gaussian components for the emission lines are plotted with colours given the in legend.}
\label{fig:spec_G}
\end{figure}
\begin{figure*}
 \includegraphics[width=\textwidth,clip]{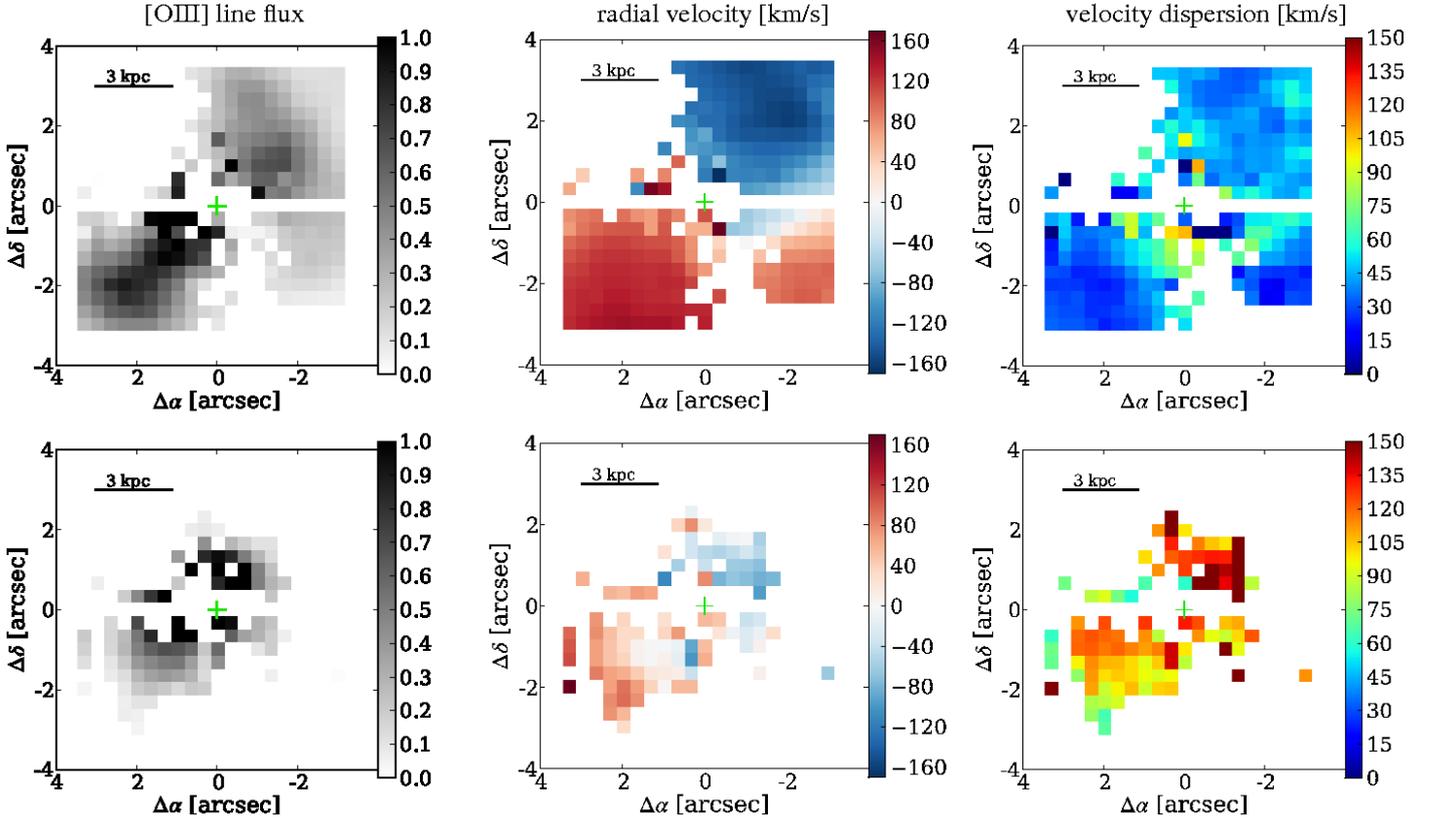}
\caption{Distribution and kinematics of two kinematically distinct emission line components. The flux distribution (left), radial velocity (middle) and velocity dispersion (right) are shown for the narrow emission line in the top panels and in the bottom panels for the broad component.}
\label{fig:kin_2comp}
\end{figure*}
\subsection{Gaseous kinematics in the inner 3\,kpc}\label{subsect:sigma_gas}
As previously mentioned in Sect.~\ref{subsect:spec_regions}, a two component model for the emission lines was required to model the spectrum of region G. We present the model for that spectrum and the individual emission-line components in Fig.~\ref{fig:spec_G}. All the strong emission lines are clearly asymmetric and a second broader Gaussian component is needed to match this asymmetry. Furthermore, we found that all emission lines share the same kinematics and were therefore kinematically coupled in the model. We inferred a velocity dispersion of 45\,km/s for the narrow and 110\,km/s for the broad component. 

From the spatially resolved kinematics assuming only a single Gaussian for the emission line, we note that the velocity dispersion is rising close to the centre. This suggests that the broad component is also spatially resolved. Thus, we modelled each spectrum of the datacube with a two-component Gaussian model for the \Ox\ doublet emission line. We employed the statistical F-test on the $\chi^2$ values for the two models to decide weather the two component model is a statistically significant better representation of the spectrum than a single component model. The resulting maps of the flux distribution, radial velocity and velocity dispersion for the narrow and broad components are shown in Fig.~\ref{fig:kin_2comp}.

These maps reveal that the narrow component originates from rotationally supported gas without a major increase in the velocity dispersion as close as 1\,kpc to the QSO. The broad component, on the other hand, is almost non-rotating with a velocity gradient of only $\pm 40$\,km/s compared to $\pm120$\,km/s of the rotating gas in the same region. The spatially resolved velocity dispersion of the gas in the broad component were measured to be in the range of $\sigma_\mathrm{g}=100-140$\,km/s in agreement with the velocity dispersion inferred from the integrated spectrum of region G. Because $\sigma/v$ is greater than 1 for the broad component, this gas seems to be dispersion dominated and mainly supported by random motion. We also detected the broad component on the blueshifted side of the rotating gas coincident with region C, which is hardly noticeable in the integrated spectrum of that region as the flux in the broad component is much lower than flux in region G. This dispersion dominated gas component in the circumnuclear region was not detectable in the previous studies by Let07 and Jah07 due to the lower spectral resolution.

\subsection{Surface brightness profiles and apparent neighbouring galaxies}\label{sect:host}

\begin{figure}
 \resizebox{\hsize}{!}{\includegraphics[clip]{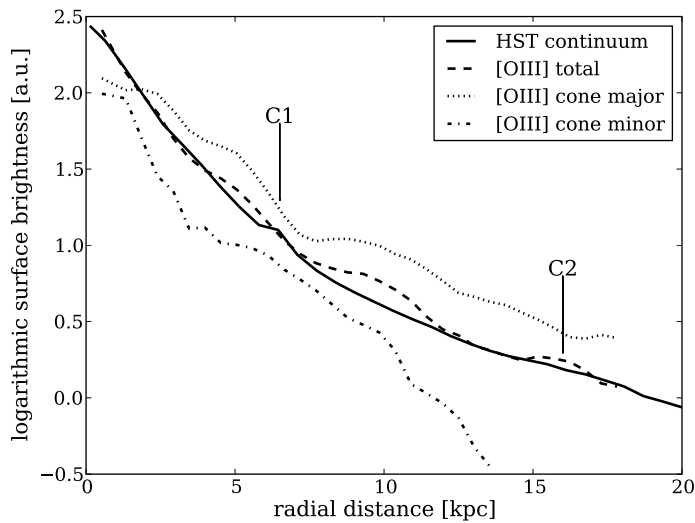}}
 \caption{Radial surface brightness profiles of the stellar continuum (solid line) compared with the total ionised gas distribution (dashed line), the ionised gas within a $70^\circ$ wide cone along the major axis of \Ox\ (dotted line) and the ionised gas for a similar cone along the minor axis (dashed-dotted line). The surface brightness profile of the stellar continuum was scaled to match that of the total ionised gas between 7.5\arcsec\ (12\,kpc) and 9\arcsec\ (14.5\,kpc) radial distance. The radial distances of the apparent companions C1 and C2 are marked for comparison.}
  \label{fig:sfb_comparison}
\end{figure}

We measured the radial surface brightness profiles of the stellar continuum emission as seen by HST and of the \Ox\ line emission from the VIMOS data using circular annuli. To account for the biconical structure of the \Ox\ light distribution we also computed surface brightness profiles within a $70^\circ$ wide cone along the major and minor axis. Due to the low spatial resolution of the VIMOS date we took the subpixel coverage of the annuli and cones into account. Since the ratio of \Ox\ to \Hb\ is constant of $\sim$10 for the entire galaxy, it is reasonable to assume that the \Ox\ light distribution is a tracer of the ionised gas distribution. The profiles are compared in Fig.~\ref{fig:sfb_comparison} after the HST surface brightness was scaled to match the total \Ox\ surface brightness in the radial distance range from 7.5\arcsec\ to 9\arcsec. 

Surprisingly, the surface brightness profile of the total ionised gas apparently follows that of the stars on a global scale despite the morphological difference. However, the profiles are significantly different for the light distribution in the cones along the major and minor axis. The drop in the \Ox\ surface brightness along the minor axis is steep with a power-law index of -1.8. This is roughly consistent with the geometrical decrease in the ionising photon density that scales as $r^{-2}$. On the other hand,  the profile along the major axis is much shallower also with respect to the stars indicating that the radial ionised gas distribution is determined by the gas density rather than the amount of available ionising photons in this direction. The distinct \Ox\ emission line knots and arcs as identified in the \Ox\ narrow-band image and channel maps  (cf. Fig.~\ref{fig:O3_image} and Fig.~\ref{fig:O3_channel}) produce excesses in the total \Ox\ surface brightness profile and along the major axis at around 5\,kpc, 10\,kpc and 16\,kpc radial distance as expected. We do not detect any corresponding features in the continuum surface brightness profile.  The stellar surface brightness profile itself is rather smooth except for the bright apparent companion C1 located 4\,arcsec away from the host centre, but we do not find any significant excess in the \Ox\ surface brightness at that location neither in the major nor the minor axis cone.

We consider two explanations for the lack of ionised gas at the location of the bright apparent neighbouring galaxy C1 north to the nucleus: 1.) The galaxy is not a physical companion so its gas cannot be ionised by the QSO. 2.) The galaxy is a physical companion with an intrinsically low amount of gas to be ionised.

The relation of the second faint galaxy to the host galaxy of HE~1029$-$1401 remains unclear. We detected weak \Ox\ emission  at its position which corresponds to one end of the large arc-like structure seen at a distance of 10\arcsec. It is possible that gas is stripped off from the faint companion while it is orbiting around the host galaxy forming a tail of gas that is partially illuminated by the QSO radiation. Since redshift information is not available for this galaxy, a simple projection effect is also a reasonable explanation.

\section{Discussion}
\subsection{The unexpected anisotropic AGN radiation field}
The biconical distribution of the ionised gas in the circumnuclear region of HE~1029$-$1401 appears to be very similar to the ionisation cones detected in several nearby Seyfert galaxies \citep[e.g.][]{Pogge:1988a,Pogge:1988b,Storchi-Bergmann:1991,Storchi-Bergmann:1992}. These collimated AGN radiation cones are supporting the unified model for AGN in which the nucleus is thought to be surrounded by a dusty torus providing a partial obscuration of the AGN. This model explains the distinction between type 1 and type 2 AGN solely by the orientation of the obscuring torus with respect to the observers line-of-sight. In this picture, the opening angle of the ionisation cones is simply determined by the covering factor of the torus.

Why is such an anisotropic radiation field on kpc scales unexpected for a luminous type 1 QSO like HE~1029$-$1401? Firstly, ionisation cones are best seen if they are oriented perpendicular to our line-of-sight, which means that the torus is oriented almost edge-on and the AGN is obscured. This is the reason why most of the known ionisation cones were detected in Seyfert 2 galaxies. Secondly, the `receding torus' model \citep{Lawrence:1991} predicts the opening angle of the AGN ionisation cones to increase with the AGN luminosity, which is observationally supported by the decreasing fraction of obscured AGN with increasing luminosity in X-ray selected AGN samples \citep[e.g.][]{Ueda:2003,Hasinger:2004,Hasinger:2008}. The X-ray 2--10\,keV luminosity of HE~1029$-$1401 is $2.3\times10^{44}\,\mathrm{erg\,s}^{-1}$ \citep[][converted to a concordance cosmology]{Reeves:2000} corresponding to a type 2 fraction of $\simeq$30\% \citep{Hasinger:2008}. From this fraction one would expect a large an opening angle of $\sim$115\degr, much more than what our observations suggest. 

\citet{Mulchaey:1996} simulated the apparent emission-line images of the ionisation cones assuming that the ambient gas distribution is a spheroid or disc. They found that for a spheroidal gas distribution the emission-line image may have a V-shape morphology but the apparent opening angle is always equal or greater than the opening angle of the ionisation cones. Due to the rotational motion of the gas, our observations indicate that the gas is more likely to be distributed in a disc rather than in a sphere. In that case, the apparent opening angle from the emission-line images is always \emph{smaller} than the opening angle of the ionisation cones depending on the angle between the disc and cone axis. If the ionisation cones is oriented such that it intercepts the disc almost at its edge, the apparent opening angle would be much smaller and consistent with our observations. Thus, the morphology of the emission-line gas around HE~1029-1401 can be consistent with the expectation of the unifed model of AGN if the ambient gas is distributed in a disc combined with a certain configuration of the AGN ionisation cones with respect to the disc. 

\subsection{The origin of the gas}
We inferred from our IFU data that the elliptical host of HE~1029$-$1401 contains ionised gas with a mass of the order of  $5\times10^6$. The ionised gas mass is rather small compared to the stellar mass of the system that has been measured to be $M_\mathrm{stellar}\approx1\times 10^{11} M_{\sun}$ based on broad-band SED modelling (M. Schramm, private communication). Unfortunately, the neutral gas fraction of HE~1029$-$1401 is unknown. Thus, it is unclear if HE~1029$-$1401 contains more gas than comparable inactive early-type galaxies, or whether the luminous QSO is just able to ionise a significant fraction of the neutral gas content.

But where did the gas come from? Our emission-line diagnostic analysis showed that the metallicity of the gas is close to solar. We now compare this with the mass-metallicity relation of star-forming galaxies as presented by \citet{Tremonti:2004}. As the mass-metallicity relation depends critically on the assumed metallicity estimator \citep{Kewley:2006}, the \citet{Tremonti:2004} relation best matches with our metallicity estimation as it is also based on the CLOUDY photoionisation code \citep{Ferland:1996}. What we find is that the expected value for the oxygen abundance for a galaxy with a stellar mass of $10^{11} M_{\sun}$ is $12+\log(\mathrm{O}/\mathrm{H})=9.11$.  This is more than 0.3\,dex larger than the solar oxygen abundance measured for HE~1029$-$1409. Although the scatter in the mass-metallicity relation is non-negligible, the metallicity of the gas in HE~1029$-$1409 would correspond to a $>$3$\sigma$ deviation from the \citet{Tremonti:2004} mass-metallicity relation at a stellar mass of $10^{11} M_{\sun}$. 

\citet{Peeples:2009} and \citet{Alonso:2010} recently showed that the high-mass low-metallicity outliners of the mass-metallicity relation are morphologically disturbed galaxies with bluer colours (or younger stellar populations) suggesting an inflow of low-metallicity gas driven by galaxy interactions and accompanied by enhanced star formation. This is consistent with the observations of HE~1029$-$1401, but the signatures for galaxy interactions are weak with only the faint shells detected by \citet{Bahcall:1997}. The high stellar mass and smooth surface brightness profile of the host suggest that the last major merger that formed the elliptical galaxy happened at a much earlier epoch. Assuming that the gas originates entirely from a single progenitor galaxy, the solar gas phase metallicity would correspond to a stellar mass of roughly $5\times 10^9M_{\sun}$ almost 10 times smaller than the host mass of HE~1029$-$1401 itself. This is consistent with the picture that ionised gas originates from one or a few minor companions that merged with HE~1029$-$1401. Therefore, we favour to interpret the low metallicity of the ionised gas in HE~1029$-$1401 as a clear indication for an \emph{external} origin of the gas via the infall of one or a few minor companions. 

Similar results have been obtained for the EELRs around the radio-loud QSOs 4C~37.43 \citep{Fu:2007} and 3C~79 \citep{Fu:2008}. The authors found in both of these cases a solar to sub-solar metallicity of the ionised gas on kpc scales around their massive elliptical host galaxies containing black holes with a mass of the order of $10^{9} M_{\sun}$. They also came to the conclusion that the metal-poor gas needs to have an external origin. 

\subsection{The kinematics of the ionised gas -- Indications for mergers or AGN outflow?}

The \Ox\ emission-line profile in the nuclear spectrum is asymmetric, which we resolved into two distinct Gaussian components with a FWHM of 421$\pm$50\,km/s and 975$\pm$129\,km/s, respectively, where the broader component is blueshifted by $120\pm50$\,km/s. Such asymmetric \Ox\ lines have been previously reported in many other QSOs \citep[e.g.][]{Heckman:1981,Veron-Cetty:2001} and were interpreted as an indication for high ionisation outflows from the nucleus \citep{Zamanov:2002}. We can not spatially resolve the extended ionised gas closer than $\sim$1\,kpc from the nucleus, but outside this unresolved region we did not find any signatures for a similarly broad and blueshifted ionised gas component indicating that this possibly outflowing component is restricted to scales below $\sim$1\,kpc.

Instead, we find that the majority of the extended ionised gas appears to be bound in a rotationally supported gas disc. The lack of rotational motion in the corresponding stellar component as inferred by Jah07 further supports an external origin of the gas from a kinematically point of view. The non-rotating dispersion-dominated ionised gas within the central 3\,kpc could either represent an \emph{intrinsic} reservoir of gas originating from the stars in the host or gas that is directly heated by the AGN. 

On larger scales the kinematics of the ionised gas around HE~1029$-$1401 strongly deviates from the rotational motion of the disc. \citet{Garcia-Lorenzo:2005} found a similarly distorted kinematic pattern in the gaseous velocity field and inner shells around the nucleus of 3C~120, which they interpreted as evidence for a merging event. Above we invoked minor mergers to explain the external origin of the gas, which would also provide a consistent explanation for the distortion in the velocity field. On the other hand, morphological distortion in the stellar component should be detected at the location of the kinematical distortions if mergers play the dominant role. It could be that the current HST image of HE~1029$-$1401 is simply too shallow to detect faint morphological distortions.  \citet{Bennert:2008} found shells and tidal tails in four out of five elliptical QSO host galaxies observed with HST in deep exposures suggesting that merger are quite common for such galaxies. But without a matched control sample of inactive elliptical galaxies this result cannot be taken as a direct evidence for a causal link between mergers and  QSO activity. Indeed, in a large sample of mainly inactive elliptical galaxies presented by \citet{Tal:2009}, similar distortions were found in 75\% of them with no correlation between AGN activity and large scale distortions. However, their AGN comparison sample for comparison consists mainly of low-luminosity radio galaxies which might have a different fuelling mechanisms compared to luminous QSOs.

 As an alternative possibility we speculate that these structures could be ionised filaments on the surface of an expanding shell/bubble driven out by the AGN. Two examples of such outflowing bubbles were investigated recently with detailed IFU observations of the Broad Absorption Line QSOs Mrk~231 \citep{Lipari:2009a} and HE~0450$-$2948 \citep{Lipari:2009b}. They argued that these outflows are part of galactic outflows driven by extreme starbursts as the result of major mergers which possibly lead also to the onset of QSO activity due to an interaction between the nuclear star formation and the central black hole. Such a particular phase of QSO activity has been claimed to be part of an evolutionary sequence of AGN \citep[e.g.][and references therein]{Sanders:1988a,Sanders:1988b,Lipari:1994b,Canalizo:2000,Papadopoulos:2008,Lipari:2006} as a short transition phase before the QSO becomes dominant. In this scenario HE~1029$-$1401 would have to be place at the end of a potential AGN evolutionary path, the QSO dominant phase. Yet, due to the unknown geometry and proper motion of these intriguing structures, it is difficult to pin down the nature of the disturbed velocity field as due to an expanding bubble. An interpretation as signatures of (minor) merging appears at least as likely.

\section{Summary and Conclusions}
We presented optical integral field spectroscopy observations of the luminous radio-quiet QSO HE~1029$-$1401 with VIMOS instrument at the VLT.  A dedicated decomposition technique for IFU data was used to decompose the spectra of the host and the QSO component. This allowed us to perform a detailed spatially-resolved analysis of the ionised gas properties based on emission-line diagnostics and a tentative analysis of the integrated stellar continuum at a S/N level of $\sim$10. We draw the following conclusions from our present study of this QSO:

\begin{itemize}
 \item We found tentative evidence from the stellar continuum for a blue featureless continuum or a young stellar population ($<$100\,Myr) embedded in the old stellar population (10\,Gyr) of the massive elliptical host galaxy. The stellar velocity dispersion is estimated to be $\sigma_*=320$\,km/s, although the uncertainty of $\pm90$\,km/s is considerable. \\

\item The  black hole mass is estimated from QSO spectrum to be $\log(M_\mathrm{BH}) = 8.7\pm0.3$ accreting at an Eddington ratio of $\log(L_\mathrm{Bol}/L_\mathrm{Edd})=-0.9\pm0.2$.  Our $M_\mathrm{BH}$ and $\sigma_*$ measurements are consistent with the $M_{\mathrm{BH}}-\sigma_*$ relation of \citet{Tremaine:2002} within the uncertainties.\\

\item We detected highly ionised gas in the host galaxy up to a projected distance of at least $\sim16$\,kpc that is solely ionised by the QSO radiation. The integrated \Ox\ luminosity of this EELR is with $\log(L(\Ox)/\mathrm{[erg/s]}) = 41.78\pm0.1$ one of the most luminous EELR detect around a radio-quiet QSO. The total ionised gas mass was estimated to $3_{-2}^{+6}\times10^6 M_{\sun}$, which should be correct at an order-of-magnitude level only.\\

\item The morphology of the ionised gas appears to be biconical with additional faint knots and arcs at larger distances. The rather narrow bicones may imply more collimated ionisation cones of the AGN than predicted by the unification model for such a luminous QSO. Assuming a disc-like distribution of gas, we prefer to interpret the biconical emission-line image as the projection of the partially illuminated gas disc.\\

\item Comparison of photoionisation models with the observed emission-line ratios of the gas indicated that the metallicity of the gas is significantly lower than expected from the mass-metallicity relation of galaxies. We interpret this as an external origin of the gas and argue that this is most likely due to minor merger(s).\\

\item We found that the ionised gas is rotationally-supported with an additional dispersion-dominated component in the circumnuclear region up to $\sim$3\,kpc. The ordered velocity field is clearly distorted on larger scales that either support the scenario of minor mergers of the gas or might be speculatively interpreted as filaments on the surface of an AGN-driven outflowing shell/bubble. However, we did not find clear and conclusive evidence for an AGN outflow.\\

\end{itemize}

Our observations draw a scenario of HE~1029$-$1401 in which the massive elliptical host has recently accreted fresh gas from its environment via minor mergers. This process leaves the host morphology almost unchanged, but the black hole accretion can be `re-juvenated' and a disc of gas is formed. If the accretion event is accompanied by a short burst of star formation, the previous red elliptical galaxy would become bluer moving the galaxy back to the `green valley' in the colour-magnitude diagram. The importance of minor mergers in the late evolution of the early-type galaxies population has recently been highlighted \citep{Naab:2009,Johansson:2009,Bezanson:2009}. It might also be an important process to explain a luminous AGN phase in early-type galaxies as indicated by our single-object study.

\begin{acknowledgements}
We thank the anonymous referee for his comments that improved the clarity of the manuscript.
BH would like to thank the members of the CEFCA institute for their hospitality during a visit were part of this work was done. BH and LW acknowledge financial support by the DFG Priority Program 1177 'Galaxy Evolution',
grant Wi 1369/22-1 and Wi 1369/22-2. SFS would like to thanks the 'Ministerio de Ciencia e Innovacion' project ICTS-2009-10, and
the 'Junta de Andalucia' projects P08-FWM-04319 and FQM360.
KJ and DN are funded through the DFG Emmy Noether-Program, grant JA 1114/3-1.

\end{acknowledgements}

\bibliographystyle{aa}
\bibliography{references}
\end{document}